\documentclass[12pt, letterpaper]{article}
\pdfoutput=1
\usepackage{amsmath,amssymb,amsfonts,cite,xcolor,epsf,epsfig,epstopdf,graphics,graphicx,mathtools,subcaption,physics,verbatim,bm}

\usepackage[utf8]{inputenc}
\usepackage{ascmac}

\def\thefootnote{*\arabic{footnote}}
\definecolor{ultramarine}{rgb}{0.07, 0.04, 0.56}
\definecolor{cadmiumgreen}{rgb}{0.0, 0.42, 0.24}
\definecolor{indigo(dye)}{rgb}{0.0, 0.25, 0.42}
\usepackage[linktocpage=true,breaklinks]{hyperref}
\hypersetup{
	colorlinks=true,
	citecolor=ultramarine,
	linkcolor=cadmiumgreen,
	urlcolor=indigo(dye),
}

\numberwithin{equation}{section}

\usepackage{array}
\newcolumntype{P}[1]{>{\centering\arraybackslash}p{#1}}

\usepackage{dcolumn}
\newcolumntype{M}[1]{>{\centering\arraybackslash}m{#1}}
\newcolumntype{N}{@{}m{0pt}@{}}

\usepackage{slashed}

\newcommand{\Mpl}{M_{\rm Pl} }
\newcommand{\diff}{ {\it Diff}_{\rm diag}(3) }
\newcommand{\Vdiff}{ {\it VDiff}_{\rm diag}(3) }

\newcommand{\expansion}{ \mathcal{K} }
\newcommand{\spatialR}{ {}^{(3)}\!\mathcal{R} }
\newcommand{\spatialgam}{ {}^{(3)}\! \gamma }
\newcommand{\spatialD}{ \mathcal{D} }
\newcommand{\spatialB}{ \mathcal{B} }
\newcommand{\spatiald}{ {\tilde{\partial}} }
\newcommand{\dertau}{ {\tilde{\partial}}_{0} }
\newcommand{\spatialV}{ {V_{\perp}} }
\newcommand{\spatialU}{ {W^{\perp}} }
\newcommand{\spatialS}{ S }
\newcommand{\spatialX}{ {X_{\perp}} }
\newcommand{\spatialY}{ {Y_{\perp}} }

\newcommand{\divh}{ (\tilde{\partial} \! \cdot \! \delta \hat{h}) }

\newcommand{\spatialRold}{ {}^{(3)}\! \widetilde{\mathcal{R}} }

\newcommand{\Lae}{\mathcal{L}_{\text{\ae}}}
\newcommand{\Lamae}{\Lambda_{\text{\ae}}}
\newcommand{\mugh}{\mu_{\text{gh}}}
\newcommand{\cgh}{c_{\text{gh},X}}

\newcommand{\nn}{\nonumber \\}

\newcommand{\D}{{\rm d}}

\newcommand{\be}{\begin{equation}}  
\newcommand{\ee}{\end{equation}}

\usepackage{paralist}

\begin{document}

%%%%%%%%%%%%%%%%%%%%%%%%%%%%%%%%%%%%%%%%%%%%%%%%%%%%%
\begin{flushright} {\footnotesize YITP-22-38\\
IPMU22-0017}  \end{flushright}
\vspace{0.5cm}

\begin{center}

\def\thefootnote{\fnsymbol{footnote}}

{\Large {\bf Effective Field Theory of Gravitating Continuum:\\
Solids, Fluids, and Aether Unified}}
\\[1cm]

{Katsuki Aoki$^{1}$, Mohammad Ali Gorji$^{1}$,
Shinji Mukohyama$^{1,2}$, Kazufumi Takahashi$^{1}$}
\\[.7cm]

{\small \textit{$^1$Center for Gravitational Physics and Quantum Information, Yukawa Institute for Theoretical Physics, Kyoto University, 606-8502, Kyoto, Japan
}}\\

{\small \textit{$^2$Kavli Institute for the Physics and Mathematics of the Universe (WPI), The University of Tokyo, 277-8583, Chiba, Japan}}

\end{center}

\vspace{.8cm}

\hrule \vspace{0.3cm}

\begin{abstract} 
We investigate the relativistic effective field theory (EFT) describing a non-dissipative gravitating continuum. In addition to ordinary continua, namely solids and fluids, we find an extraordinary more symmetric continuum, aether. In particular, the symmetry of the aether concludes that a homogeneous and isotropic state behaves like a cosmological constant. We formulate the EFT in the unitary/comoving gauge in which the dynamical degrees of freedom of the continuum (phonons) are eaten by the spacetime metric. This gauge choice, which is interpreted as the Lagrangian description in hydrodynamics, offers a neat geometrical understanding of continua. We examine a \emph{thread-based} spacetime decomposition with respect to the four-velocity of the continuum which is different from the foliation-based Arnowitt-Deser-Misner one. Our thread-based decomposition respects the symmetries of the continua and, therefore, makes it possible to systematically find invariant building blocks of the EFT for each continuum even at higher orders in the derivative expansion. We also discuss the linear dynamics of the system and show that both gravitons and phonons acquire ``masses'' in a gravitating background.
\end{abstract}
\vspace{0.5cm} 

\hrule
\def\thefootnote{\arabic{footnote}}
\setcounter{footnote}{0}

\thispagestyle{empty}

\newpage
\hrule
\tableofcontents
%\addtocontents{toc}{\protect\setcounter{tocdepth}{2}} 
\vspace{0.7cm}
\hrule

\newpage
%%%%%%%%%%%%%%%%%%%%%%%%%%%%%%%%%%%%%%%%%%%%%%%%%%%%%%%%%%%%%%%%%%%%%%%%%%%%%%%%%%%%
%%%%%%%%%%%%%%%%%%%%%%%%%%%%%%%%%%%%%%%%%%%%%%%%%%%%%%%%%%%%%%%%%%%%%%%%%%%%%%%%%%%%
%	Introduction
%%%%%%%%%%%%%%%%%%%%%%%%%%%%%%%%%%%%%%%%%%%%%%%%%%%%%%%%%%%%%%%%%%%%%%%%%%%%%%%%%%%%
%%%%%%%%%%%%%%%%%%%%%%%%%%%%%%%%%%%%%%%%%%%%%%%%%%%%%%%%%%%%%%%%%%%%%%%%%%%%%%%%%%%%
\section{Introduction and summary}\label{introduction}
Continuum is ubiquitous in nature. Although different materials have different microscopic properties, they can be modeled to be distributed continuously at a macroscopic level and their macroscopic properties may have similarities. Continuum mechanics deals with such continuously distributed materials including solids and fluids, and the concept of continuum appears in many areas of physics. In recent years, continuum mechanics, and especially one branch of it, hydrodynamics, have been reformulated in a modern language, that is, low-energy effective field theory~(EFT)~\cite{Dubovsky:2005xd,Endlich:2010hf,Dubovsky:2011sj,Endlich:2012pz,Kovtun:2014hpa,Haehl:2015pja,Harder:2015nxa,Crossley:2015evo,Haehl:2015uoc,Jensen:2017kzi,Glorioso:2017fpd,Haehl:2018lcu}. The EFT approach not only reformulates continuum mechanics in a modern way using the action principle instead of equations of motion but also reveals the underlying first principle, i.e.~symmetries, and then provides a systematic way to investigate microscopic corrections as characterized by higher derivative operators. However, gravitational effects have not been well investigated in the literature. In the present paper, we would like to highlight roles of gravity and systematically develop the EFT of {\it gravitating} continuum describing general-relativistic continuum mechanics.

The importance of gravity is two-fold. First of all, gravity plays essential roles in relativistic phenomena such as astrophysics and cosmology in which matter fields are usually treated as fluids. The increasing amount of astrophysical/cosmological observations including gravitational waves have been leading to a better understanding of gravity and the matter fields in such phenomena and, therefore, it is desirable to systematically develop the underlying theories. Secondly, dynamical gravity is necessary to understand the spontaneous symmetry breaking of spacetime symmetries caused by the presence of a continuum. As we will explain below, the gauge symmetry of gravity, namely the spacetime diffeomorphisms~(diffs), is spontaneously broken when a spacetime is filled with a continuum. 

It is apparently straightforward to incorporate gravity: one may first consider an EFT in the flat spacetime and then replace the flat spacetime metric and the partial derivatives by the curved spacetime metric and the covariant derivatives, respectively. In this conventional approach, in order to make such replacement unambiguously one may need to start with the Lorentz-invariant description of the theory although the system itself has no Lorentz invariance in the broken phase. For instance, in the case of solids, one first needs to separately deal with the spacetime $ISO(3)$ symmetry and the internal $ISO(3)$ symmetry~\cite{Endlich:2012pz}. The former is a part of the full spacetime symmetry while the latter is required to describe a continuum with desired properties as we will explain later. On the other hand, we are interested in the symmetry breaking pattern $ISO(3)\times ISO(3) \to ISO_{\rm diag}(3)$, where $ISO_{\rm diag}(3)$ is the unbroken diagonal part of the $ISO(3)$ symmetries, which is caused by expectation values of first-order derivatives of fields. Hence, the conventional approach requires 
\begin{inparaenum}[(i)]
\item \label{step1} writing down the most general action with the spacetime covariance and the internal $ISO(3)$ invariance while treating the first-order derivatives of the fields non-perturbatively and 
\item \label{step2} reorganizing the action in terms of the perturbations around the broken phase.
\end{inparaenum}
On the other hand, bypassing the two steps~\eqref{step1} and \eqref{step2}, it is possible to directly write down the EFT action which only respects the unbroken $ISO_{\rm diag}(3)$ symmetry, namely, the theory of perturbations around the broken phase, and straightforwardly perform the derivative expansion, though coupling to gravity requires a careful treatment due to the lack of the spacetime covariance in this case. In the present paper, we reformulate the EFT based on this latter approach.

In cosmology, the observed isotropy and homogeneity at large scales fix the background geometry of the universe to be the Friedmann‐Lema\^itre‐Robertson-Walker (FLRW) metric. There are many candidates for the matter configurations which are compatible with the FLRW background and the following question may arise: which one is preferred by our universe? To answer this question, one needs to study cosmological perturbations which characterize small deviations from the perfect isotropy and homogeneity. Perturbations produced by different matter fields have different statistical properties and, therefore, different matter fields may be distinguished through the cosmological observations. In this regard, it is quite important to study perturbations produced by different matter fields. The EFTs of cosmology provide appropriate frameworks to study cosmological perturbations around the FLRW background in a model-independent way. The simplest matter configuration is a single scalar field for which the EFT setup is well known~\cite{Creminelli:2006xe,Cheung:2007st,Creminelli:2008wc,Gubitosi:2012hu,Bloomfield:2012ff,Gleyzes:2013ooa,Gleyzes:2014rba}. On the other hand, as we mentioned, there are other candidates like solids with which we are interested in this paper \cite{Endlich:2012pz}. Other examples are superfluids, supersolids, framids~\cite{Nicolis:2015sra} and gauge fields~\cite{Maleknejad:2011jw,Maleknejad:2012fw}. The symmetry breaking pattern for the solids is different from the one in the so-called EFT of inflation/dark energy~\cite{Creminelli:2006xe,Cheung:2007st,Creminelli:2008wc,Gubitosi:2012hu,Bloomfield:2012ff,Gleyzes:2013ooa,Gleyzes:2014rba} so that standard techniques such as the Arnowitt-Deser-Misner (ADM) decomposition are not useful anymore. Thus, in the present paper, we will elaborate on how the spacetime is decomposed into space and time to preserve the symmetries required by the solids and identify invariant geometrical quantities and a convenient gauge choice to write down the EFT for solids. It will turn out that the resultant EFT action for the solid inflation is quite analogous to that of the EFT of single-field inflation~\cite{Cheung:2007st} but with a clear difference due to the difference in the underlying symmetry breaking patterns. 

Thanks to employing the spacetime decomposition proper for the continua, our EFT formulation not only includes solids but also includes fluids and, more interestingly, reveals a novel phase of continuum, \emph{aether}, which is a quite symmetric object and shows peculiar behaviors due to its symmetry. In particular, the aether behaves like a cosmological constant at the background level when the distribution is homogeneous and isotropic and it may be a natural candidate for dark energy. Thus, our formulation provides a unified framework for the EFT of solids, fluids, and aether which can be described by the same variables but have different internal symmetries. The types of the continua are distinguished by absence/existence of EFT operators belonging to the different symmetry groups of the broken phase. If possible, it would be interesting to generalize our formulation to unify superfluids, supersolids, and framids as the cosmological perturbation theory of the scalar-tensor theories (superfluids coupled to gravity) and that of the vector-tensor theories (type-I framids coupled to gravity) have been unified into a single framework~\cite{Aoki:2021wew}. Moreover, the new spacetime decomposition allows us to describe the EFT of solids in the unitary/comoving gauge, while the solid inflation has been well understood only in the spatially flat gauge~\cite{Endlich:2012pz}. Although this may seem as just rewriting the same theory in a different gauge, it will be very important at least for the following three reasons. First, the reheating process for the solid inflation is not understood yet and working in the unitary gauge may be considered as a first step in this direction. Secondly, working in the unitary gauge makes it possible to systematically include higher-order operators. This becomes very important when the higher-order operators become dominant due to the high symmetry of the system under consideration. For instance, this is the case for the aether as we will show. Thirdly, as it is well known, the consistency conditions among correlation functions for the standard single-field inflation are violated in solid inflation~\cite{Endlich:2013jia,Dimastrogiovanni:2014ina,Akhshik:2014bla,Bordin:2016ruc} and our EFT provides an appropriate framework to compute higher-order correlation functions.

For the convenience of the readers, we summarize below the conventions and main results of the present paper with the reference to sections in which the readers can find details. Let us also summarize several terminologies since different terms would be more familiar to different fields.

\begin{itemize}
\item Our interest is a long-range/time dynamics of a gravitating material which can be treated as a continuum in a dynamical spacetime of four dimensions. The continuum has its preferred spatial coordinates, i.e., {\it Lagrangian (or comoving) coordinates}, which are comoving with individual fluid particles. The Lagrangian coordinates are denoted by $\phi^i~(i=1,2,3)$ and the mapping between the Lagrangian coordinates and spacetime (Eulerian) coordinates are given by $\phi^i=\phi^i(t,\bm{x})$ where $\bm{x}$ denotes the spatial coordinates. Hence, our essential interest is the dynamics of the triplet of scalar fields~$\phi^i(t,\bm{x})$ as well as the spacetime metric~$g_{\mu\nu}(t,\bm{x})$. See Sec.~\ref{sec:solidetc}.

\item The background configuration of the fields~$\phi^i$ is characterized by the vacuum expectation values
\begin{align}\label{BG}
&\langle \phi^i \rangle = x^i \,,
&\mbox{({\it background configuration})}
\end{align} 
which spontaneously breaks parts of the spacetime symmetries. Throughout, we assume that the background is homogeneous and isotropic, which is guaranteed by the internal symmetries of the fields~$\phi^i$.
Although the background configuration is similar to the cosmological EFTs~\cite{Creminelli:2006xe,Cheung:2007st,Creminelli:2008wc,Gubitosi:2012hu,Bloomfield:2012ff,Gleyzes:2013ooa,Gleyzes:2014rba,Abolhasani:2015cve,Rostami:2017wiy,Gong:2019hwj,Aoki:2021wew}, the different symmetry breaking pattern will lead to a physically different scenario when we take into account perturbations \cite{Endlich:2012pz}. See Sec.~\ref{sec:solidetc}.

\item Let us refer to the unitary/comoving gauge in which the spatial coordinates agree with the Lagrangian coordinates even in the presence of the perturbations about the background~\eqref{BG},
\begin{align}\label{unitary-gauge}
&\phi^i=x^i \,.
&\mbox{({\it unitary/comoving gauge})}
\end{align}
The dynamical spacetime metric~$g_{\mu\nu}(t,\bm{x})$ plays the role of a gauge field that eats the degrees of freedom of the continuum, analogously to what happens in a spontaneously broken gauge theory. On the other hand, this gauge is also {\it the comoving gauge} from the hydrodynamical perspective and the resultant description of the continuum is referred to as {\it the Lagrangian description}. See Secs.~\ref{sec:solidetc} and \ref{sec:hydro_decomposition}.

\item These types of continua are classified by their response to a time-independent deformation and we will consider three types of materials accordingly: {\it solids}, {\it fluids}, and {\it aether}. The solids are materials that possess a stress against a deformation; that is, the solids are not invariant under deformations. The fluids are, on the other hand, invariant under deformations without compression or dilatation. Finally, the aether is a more symmetric object, invariant under any type of time-independent deformations. The different types lead to different residual symmetries of the broken phase. See Secs.~\ref{sec:solidetc} and \ref{sec:building_blocks}.

\begin{figure}[t]
 \centering
 \includegraphics[width=0.75\linewidth]{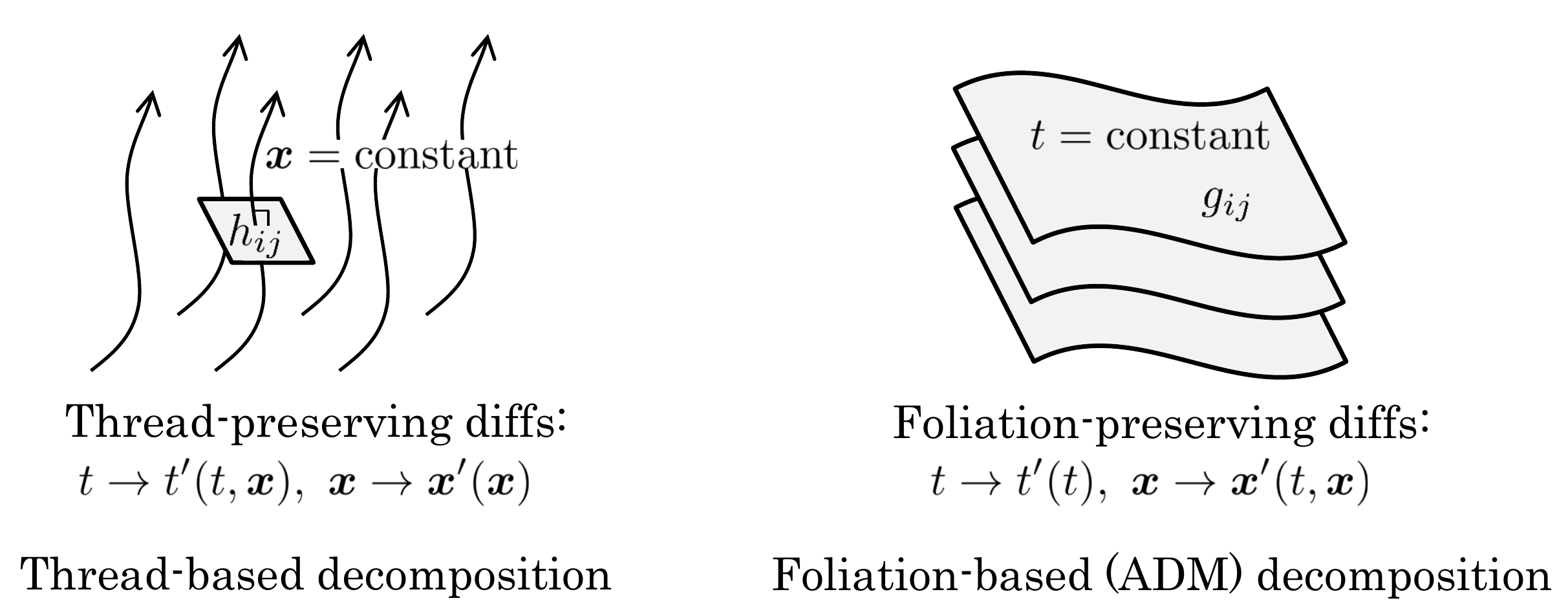}
 \caption{We decompose the spacetime into constant-position curves (threads) while the spacetime is decomposed into constant-time hypersurfaces (foliations) in the foliation-based (ADM) decomposition. The thread-based decomposition and the foliation-based decomposition preserve different parts of the spacetime diffs. In the unitary/comoving gauge, each fluid particle moves along an ${\bm x}=$ constant curve.}
 \label{fig:decomposition}
\end{figure}

\item In the unitary/comoving gauge, the only dynamical variable is the spacetime metric~$g_{\mu\nu}(t,\bm{x})$. We will elaborate later that the most convenient way to express the spacetime metric is by the use of variables~$\{U, U_i, h_{ij} \}$ as follows:
\begin{align}
g_{\mu\nu}=
\begin{pmatrix}
-U^2 & U^2 U_j \\
U^2 U_i & h_{ij}-U^2 U_i U_j
\end{pmatrix}
\,,
\end{align}
which we call {\it the thread-based decomposition} of the spacetime (Figure~\ref{fig:decomposition}). 
We also introduce several geometrical quantities as summarized in Table~\ref{table:notation}. The utility of the thread-based decomposition is two-fold. On the one hand, one can easily read hydrodynamical quantities from the spacetime metric: the kinematical quantities of the continuum, say the expansion, are computed as shown in Table~\ref{table:notation} and the variables~$\{U, U_i, h_{ij} \}$ are conjugate to the energy density~$\rho$, the flux~$q^i$ and the stress tensor~$\tau^{ij}$ in the sense that they are obtained by the variations with respect to $\{U, U_i, h_{ij} \}$: 
\begin{align}
    \rho &= -\frac{1}{\sqrt{h}} \frac{\delta S_{\rm m}}{\delta U}
    \,, \qquad
    q^i = \frac{1}{U^2 \sqrt{h}} \frac{\delta S_{\rm m}}{\delta U_i}
    \,, \qquad
    \tau^{ij} = \frac{2}{U\sqrt{h}} \frac{\delta S_{\rm m}}{\delta h_{ij}}
    \,,
\end{align}
where $S_{\rm m}$ is the action of the continuum which we will introduce below. On the other hand, the 
quantities of the thread-based decomposition have nice transformation properties under {\it the thread-preserving diffs}, $t\to t'(t,\bm{x}), \bm{x} \to \bm{x}'(\bm{x})$. The symmetry transformations of continua are recast as parts of the thread-preserving diffs in the unitary/comoving gauge. Note that, as illustrated in Figure~\ref{fig:decomposition}, the thread-based decomposition is different from the foliation-based decomposition known as ADM decomposition so that the spacetime has been decomposed with respect to constant-position curves (threads), rather than constant-time hypersurfaces (foliations). For instance, $h_{ij}$ is not a metric for a constant-time hypersurface but a metric for a tangent space orthogonal to the four-velocity. Tensors in the thread-based decomposition are spatial in the sense that they are orthogonal to the timelike threads, so we shall denote them by {\it ortho-spatial} tensors. See Secs.~\ref{sec:hydro_decomposition} and \ref{sec:building_blocks}.

\begin{table}[t]
  \caption{Summary of symbols. }
  \label{table:notation}
  \centering
  \begin{tabular}{lll}
      \hline
      Symbol & Description
      \\
      \hline \hline
      $i,j,k,\cdots $ & ortho-spatial indices (Lagrangian frame) 
      \\
      $h_{ij},h^{ij}$ & ortho-spatial metric and inverse
      \\
      $h={\rm det}h_{ij}$ & metric determinant 
      \\
      $n=h^{-1/2}$ & number density 
      \\
     $\hat{h}_{ij}=h^{-1/3}h_{ij},~\hat{h}^{ij}=h^{1/3}h^{ij}$ & unimodular metric and inverse
      \\
      $\delta \hat{h}_{ij}=\hat{h}_{ij}-\delta_{ij}$ & perturbation of unimodular metric
      \\
       $\dertau = \frac{1}{U}\partial_0 $ & material derivative 
      \\
       $\spatiald_i = \partial_i + U_i \partial_0 $ & ortho-spatial derivative
      \\
      $\expansion_{ij}=\frac{1}{2}\dertau h_{ij},~\expansion=\expansion^i{}_i$ & expansion tensor and scalar  
      \\
      $\sigma_{ij}=\expansion_{ij}-\frac{1}{3}\expansion h_{ij}$ & shear tensor 
      \\
      $a_i = U( \dertau U_i - \spatiald_i U^{-1}) $  & acceleration vector 
      \\
     $\omega_{ij} = U \spatiald_{[i} U_{j]}$ & vorticity tensor 
      \\
      $\spatialD_i$ \quad [See Eq.~\eqref{eqs:DV}] & ortho-spatial covariant derivative
      \\
      $\spatialgam^i{}_{jk} = \frac{1}{2}h^{il}\left( \spatiald_{j} h_{kl} + \spatiald_k h_{jl} -\spatiald_l h_{jk} \right)$ & connection compatible with $h_{ij}$ 
      \\
      $\spatialR^i{}_{jkl}= 2\left( \spatiald_{[k}\spatialgam^i{}_{j|l]}+ \spatialgam^i{}_{n[k}\spatialgam^n{}_{j|l]} \right)$ & ortho-spatial curvature 
      \\
      $\spatialR_{ij}=\spatialR^k{}_{ikj},~ \spatialR=\spatialR^i{}_i$ & ortho-spatial Ricci tensor and scalar 
      \\
      \hline
  \end{tabular}
\end{table}

\item Around the gravitating homogeneous and isotropic background with the scale factor~$a(t)$ which is time-dependent due to the gravitational attraction, we can choose {\it the uniform number density slice}, $n=n(t)=a^{-3}(t)$, consistently with the unitary/comoving gauge conditions~\eqref{unitary-gauge}. In this gauge, all the degrees of freedom are encoded in the spacetime metric and the leading-order action, up to the quadratic order in perturbations and derivatives, is given by
\begin{align}
S[g_{\mu\nu}]&= \int \D t \D^3 {\bm x} U(t,\bm{x}) a^3(t) \Bigl( \mathcal{L}_{\rm EH}[g_{\mu\nu}] + \mathcal{L}_{\rm m}[g_{\mu\nu}] \Bigr)
\,, \label{intro_action}
\\
\mathcal{L}_{\rm EH}[g_{\mu\nu}]&= \frac{\Mpl^2}{2}R[g_{\mu\nu}] = \frac{\Mpl^2}{2}\left[ \spatialR + \expansion_{ij} \expansion^{ij}-\expansion^2  +\omega_{ij}\omega^{ij} \right] + \text{boundary terms}
\,, \\
\mathcal{L}_{\rm m}[g_{\mu\nu}]&= \rho_* \left[ F_0(t) + F_{e_2}(t) e_2+ \frac{1}{2\Lamae^2}\left( b_{\expansion^2} \expansion^2 + b_{\sigma^2}\sigma_{ij}\sigma^{ij} + b_{\omega^2} \omega_{ij}\omega^{ij} +b_{a^2} a_i a^i \right) \right]
,
\end{align}
where $R[g_{\mu\nu}]$ is the four-dimensional Ricci scalar and $e_2 \equiv  \frac{1}{2} [ (\hat{h}^{ij}\delta \hat{h}_{ij})^2-  \hat{h}^{ij}\hat{h}^{kl}\delta \hat{h}_{jk} \delta \hat{h}_{li} ]$ is the degree-$2$ elementary symmetric polynomial of the eigenvalues of $\hat{h}^{ik}\delta\hat{h}_{kj}$.\footnote{The action is organized in a way similar to cosmological EFTs~\cite{Creminelli:2006xe,Cheung:2007st,Creminelli:2008wc,Gubitosi:2012hu,Bloomfield:2012ff,Gleyzes:2013ooa,Gleyzes:2014rba,Abolhasani:2015cve,Rostami:2017wiy,Gong:2019hwj,Aoki:2021wew}, but there is a small difference. See \ref{sec:gravitatingBG} and the footnote~\ref{footnote:expansion}.}
Here, $\rho_*$ is a constant with the dimension of energy density and $\Lamae$ is a cutoff scale. The properties of the continua are determined by the dimensionless functions~$\{ F_0(t), F_{e_2}(t) \}$ and the dimensionless constants~$\{ b_{\expansion^2},b_{\sigma^2},b_{\omega^2},b_{a^2} \}$. The solids correspond to $F_{e_2} \neq 0$ and $F_0 \neq 0$ while the fluids correspond to the case $F_{e_2} =0$ with $F_0 \neq 0$. In the case of solids/fluids, the dynamics of the continuum is governed by the lowest-derivative operators~$F_0$ and/or $F_{e_2}$ and then the terms suppressed by $\Lamae$ can be negligible at low energy. In particular, the action of the perfect fluid, which is defined as the fluid without higher-derivative corrections, is solely determined by $F_0(t)$ even at nonlinear orders of perturbations. However, the symmetry of the aether requires $F_0= $ constant and $F_{e_2}=0$, meaning that the lowest-derivative operator is a cosmological constant. Hence, the operators suppressed by $\Lamae$ become the leading ones for the dynamics of the aether. See Secs.~\ref{subsec:leading} and \ref{sec:gravitatingBG}.

\item Let us explain how the action~\eqref{intro_action} has bypassed the two steps~\eqref{step1} and \eqref{step2} required in the conventional approach mentioned above. Recall that, in the conventional approach, \eqref{step1} we should first write down the most general action under the general covariance and the internal symmetry,
\begin{align}
    S[g_{\mu\nu},\phi^i] = \int \D^4 x \sqrt{-g}  \mathcal{L} (g_{\mu\nu},R_{\mu\nu\rho\sigma},\phi^i,\nabla_{\mu}) 
    \,,
    \label{intro_action2}
\end{align}
where $R_{\mu\nu\rho\sigma}$ and $\nabla_{\mu}$ are the spacetime curvature and the spacetime covariant derivative, respectively. Here, the spacetime indices~$\mu,\nu,\cdots$, their spatial components, and the internal indices~$i,j,\cdots$ should be distinguished. In the case of solids, the internal $ISO(3)$ symmetry requires that $\phi^i$ must be differentiated at least once and the indices~$i,j,\cdots$ have to be contracted by $\delta_{ij}$. Although the invariant action takes a tractable form at the leading order in the derivative expansion~\cite{Endlich:2012pz}, the action becomes highly complicated when higher-order derivative corrections are taken into account (see, e.g.,\cite{Sivanesan:2013tba,Allys:2016hfl,Aoki:2021kla} for higher-derivative interactions under the restriction that there is no Ostrogradsky ghost). Furthermore, for practical calculations, \eqref{step2} one should split the variables into the background and the perturbations and then reorganize the action~\eqref{intro_action2} to the power of perturbations. On the other hand, since our EFT action~\eqref{intro_action} only involves ortho-spatial indices~$i,j,\cdots$ thanks to using the unitary gauge and the thread-based decomposition, it is easier to identify  invariant operators even at a higher order in derivatives [see \eqref{intro_action} and Appendix~\ref{app:HDO}]. In addition, the building blocks of the EFT are classified with respect to the orders of perturbations; for example, $e_2$ starts at quadratic order in perturbations. Hence, one can easily recognize at which order of perturbations each operator starts to contribute; $e_2$ starts at quadratic order, $e_2^2$ starts at quartic order [that is why $e_2^2$ is not included in the quadratic order action~\eqref{intro_action}], and so on. If one prefers the spacetime covariant action~\eqref{intro_action2}, one can use the St\"{u}ckelberg trick to restore the general covariance from our EFT action~\eqref{intro_action}. From the hydrodynamical perspective, this is a transformation from the Lagrangian coordinates to the Eulerian coordinates. See Secs.~\ref{sec:Eulerian}, \ref{subsec:leading}, and \ref{sec:gravitatingBG}.

\item The Nambu-Goldstone bosons associated with the present symmetry breaking pattern can be interpreted as {\it the phonons}, which appear as the {\it transverse} and {\it longitudinal} modes of {\it the graviton} in the unitary/comoving gauge. There are also the {\it transverse-traceless} modes representing the usual polarization modes of the graviton. The transverse-traceless modes, the transverse modes, and the longitudinal modes are often called {\it the tensor modes} (T), {\it the vector modes} (V), and {\it the scalar modes} (S), respectively, especially in the context of cosmology. We do not use L and T for longitudinal and transverse modes to avoid overlapping with the tensor mode (T). See Secs.~\ref{sec:Eulerian}, \ref{subsec:decoupling}, and \ref{sec:perturbations}.

\item The dynamical properties of each mode depend on the types of continua, stemming from the residual symmetries of the broken phase. All modes acquire ``masses'' in the solid phase and the masses are proportional to the sound speeds of different propagation states. In the fluid phase, the transverse phonon is non-propagating (zero sound speed) and the transverse-traceless graviton and the longitudinal phonon become massless. Not only the transverse phonons but also the longitudinal phonon become non-propagating in the aether phase. However, the aether possesses a peculiar property and it can admit other propagating longitudinal and transverse modes when the symmetry of the aether is exact. These propagating modes can be identified with those of the Einstein-aether theory~\cite{Jacobson:2007veq}. See Secs.~\ref{subsec:Noether}, \ref{subsec:decoupling}, and \ref{sec:perturbations}.

\end{itemize}

The rest of the paper is organized as follows. Section~\ref{sec:generalities} is devoted to general discussions of the EFT without specifying a concrete form of the action. We then determine the action by employing the derivative expansion and discuss linear dynamics under the decoupling limit of gravity in Sec.~\ref{sec:derivative}. In Sec.~\ref{sec:gravitating_continuum}, including dynamical gravity and fixing the gauge, we show that the EFT action is given by the simple form~\eqref{intro_action}. 
We also clarify the geometrical meaning of perturbation variables and study gravitational perturbations in terms of the geometrical quantities. Future directions are discussed in Sec.~\ref{sec:discussion}.

%%%%%%%%%%%%%%%%%%%%%%%%%%%%%%%%%%%%%%%%%%%%%%%%%%%%%%%%%%%%%%%%%%%%%%%%%%%%%%%%%%%%
%%%%%%%%%%%%%%%%%%%%%%%%%%%%%%%%%%%%%%%%%%%%%%%%%%%%%%%%%%%%%%%%%%%%%%%%%%%%%%%%%%%%
%	Generalities
%%%%%%%%%%%%%%%%%%%%%%%%%%%%%%%%%%%%%%%%%%%%%%%%%%%%%%%%%%%%%%%%%%%%%%%%%%%%%%%%%%%%
%%%%%%%%%%%%%%%%%%%%%%%%%%%%%%%%%%%%%%%%%%%%%%%%%%%%%%%%%%%%%%%%%%%%%%%%%%%%%%%%%%%%
\section{Generalities}
\label{sec:generalities}

%%%%%%%%%%%%%%%%%%%%%%%%%%%%%%%%%%%%%%%%%%%%%%%%%%%%%%%%%%%%%%%%%%%%%%%%%%%%%%%%%%%%
%	Symmetry breaking pattern
%%%%%%%%%%%%%%%%%%%%%%%%%%%%%%%%%%%%%%%%%%%%%%%%%%%%%%%%%%%%%%%%%%%%%%%%%%%%%%%%%%%%
\subsection{Solids, fluids, aether}
\label{sec:solidetc}
Throughout the present paper, we consider gravitating non-dissipative continuous materials. We first clarify the field content and symmetries to describe this system by following \cite{Dubovsky:2005xd,Endlich:2010hf,Dubovsky:2011sj,Endlich:2012pz}. On top of the ordinary continuous materials, namely solids and fluids, we will also introduce another continuum, aether, which can be understood as a continuum with an extended symmetry group than the ordinary materials. A vector field is usually used to describe aether while aether from the perspective of continuum has not been reported to the best of our knowledge.

Let $\phi^i~(i=1,2,3)$ be the Lagrangian (comoving) coordinates of the fluid particles of the continuum at a time~$t$. The mapping between the Lagrangian coordinates and the spacetime coordinates are given by a triplet of spacetime scalars~$\phi^i=\phi^i(t,\bm{x})$ where $\bm{x}$ denotes spatial coordinates of the physical spacetime. Since the system gravitates, the physical spacetime is supposed to be dynamical and its metric tensor is denoted by $g_{\mu\nu}=g_{\mu\nu}(t,\bm{x})$, where $\mu, \nu = 0, 1, 2, 3$. The set~$\{g_{\mu\nu}(t,\bm{x}),\phi^i(t,\bm{x})\}$ represents the basic dynamical variables of the present system.

The full theory respects invariance under the spacetime diffs,
\begin{align}
t \to t'=t'(t,\bm{x})\,, \qquad \bm{x} \to \bm{x}'=\bm{x}'(t,\bm{x})
\,,
\label{spacetime_diffs}
\end{align}
and some internal symmetries of $\phi^i$ specifying properties of the continuum which we will mention below. On the other hand, we are interested in the spacetime filled with the continuum and then the spacetime symmetries are spontaneously broken. We assume that the background of the continuum corresponds to
\begin{align}
\langle \phi^i \rangle = x^i
\,. \label{background}
\end{align}
Let us then suppose that the continuum has no particular positions and directions which is realized by imposing the global $ISO(3)$ symmetry:
\begin{align}\label{ISO3-phi}
\phi^i \to \phi^i + c^i\,, \qquad \phi^i \to O^i{}_j \phi^j \,,
\end{align}
where $c^i=$ constant and $O^i{}_j$ is a constant orthogonal matrix with unit determinant satisfying $O^i{}_kO^k{}_j=\delta^i_j$. 
This symmetry characterizes the (isotropic) solid.
The existence of the $ISO(3)$ symmetry concludes that the background is homogeneous and isotropic. More precisely, the background is invariant under the diagonal part of the spacetime and internal $ISO(3)$ transformations. As a subset of \eqref{spacetime_diffs}, the setup includes the spatial translation and rotation symmetry group~$ISO(3)$, 
\begin{align}\label{ISO3-x}
x^i \to x^i + d^i\,, \qquad x^i \to R^i{}_j x^j \,,
\end{align}
where $d^i$ are constant parameters and $R^i{}_j$ is a constant orthogonal matrix with unit determinant. The full theory then enjoys $ISO(3)\times{ISO(3)}$ symmetry. However, the background \eqref{background} is invariant only under the subset~$d^i=c^i$ and $R^i{}_j=O^i{}_j$ which is given by
\begin{equation}
\begin{split}
x^i &\to x^i + c^i\,, \qquad \phi^i \to \phi^i + c^i \,, \\
x^i &\to O^i{}_j x^j \,, \qquad \phi^i \to O^i{}_j \phi^j
\,.
\label{ISOdiag}
\end{split}
\end{equation}
Therefore, the symmetry breaking pattern is represented by $ISO(3)\times{ISO(3)}\to{ISO}_{\rm diag}(3)$, where ${ISO}_{\rm diag}(3)$ denotes the diagonal part of the spacetime \eqref{ISO3-x} and internal \eqref{ISO3-phi} $ISO(3)$ symmetries.

The fluid phase is distinguished from the solid phase by reaction under deformations of the material without compression or dilation. The solid possesses a stress against displacements of the volume elements while the fluid can be deformed without a stress so long as the deformation is non-compressive or non-dilative. The fluid is thus described by enlarging the internal symmetry into the volume-preserving diffs,
\begin{align}
\phi^i \to \phi'{}^i(\phi^j) \quad {\rm s.t.} \quad {\rm det}\frac{\partial \phi'{}^i}{\partial \phi^j} = 1
\,,
\label{Vdiffs}
\end{align}
which discriminates the fluid from the (isotropic) solid that only possesses the internal $ISO(3)$ symmetry~\eqref{ISO3-phi}.

We can also introduce a third possible phase of the continuum, aether, which is characterized by the invariance under the generic (not necessarily volume-preserving) internal diffs,
\begin{align}
\phi^i \to \phi'{}^i(\phi^j) \,.
\label{generaldiffs}
\end{align}
The reason to call it aether is clarified by inspecting the four-velocity.
The four-velocity of continuum is defined by
\begin{align}
u^{\mu}&\equiv -\frac{1}{6 \sqrt{{\rm det} h^{mn}} } \varepsilon_{ijk}\epsilon^{\mu\nu\rho\sigma}\partial_{\nu}\phi^i \partial_{\rho}\phi^j \partial_{\sigma} \phi^k
\,, 
\end{align}
which arises as a solution to
\begin{align}
u^{\mu}u_{\mu}=-1
\,, \qquad u^{\mu}\partial_{\mu}\phi^i = 0\,,
\end{align}
where $h^{ij} \equiv g^{\mu\nu}\partial_{\mu}\phi^i \partial_{\nu} \phi^j$. The first condition is the normalization of $u^{\mu}$ and the second condition states that the Lagrangian coordinates must be unchanged along the flow. Here, $\epsilon^{\mu\nu\rho\sigma}$ and $\varepsilon_{ijk}$ are the spacetime Levi-Civita tensor and the anti-symmetric symbol with $\epsilon^{0123}=-1/\sqrt{-g}$ and $\varepsilon_{123}=1$, respectively. Note that $u^{\mu}$ is a spacetime vector and invariant under the generic internal diffs. Therefore, in the aether phase we cannot use $\phi^i$ but can use $g_{\mu\nu}$ and $u^{\mu}$ as the building blocks of the EFT. The aether has no proper notion of the volume since the volume is no longer invariant under \eqref{generaldiffs}. The energy density of the aether must be independent of the absolute value of the volume, implying that the background behaves as a cosmological constant. Still, the aether has a notion of the flow described by the unit timelike vector $u^{\mu}$ which determines the preferred direction of the spacetime. These are exactly the behaviors of the aether and indeed the Einstein-aether theory~\cite{Jacobson:2007veq} arises as a special case of our aether continuum. See Appendix~\ref{app:Einsteinaether} for details.

As we stated, we are focusing on the symmetry breaking phase with the background~\eqref{background}, and hence fluctuations around the background are given by $\delta \phi^i(t,{\bm x}) \equiv \phi^i(t,{\bm x}) - x^i$. Let us consider small fluctuations and discuss the transformations of $\delta \phi^i$ under the infinitesimal spacetime~\eqref{spacetime_diffs} and internal diffs~\eqref{Vdiffs}. The fluctuations~$\delta \phi^i$ transform as scalars with respect to the time diffs while they transform nonlinearly under the spatial diffs:
\begin{align}
t &\to t+\xi^{0}(t,\bm{x})\,,\qquad x^i\to x^i\,, \qquad \delta \phi^i(t, \bm{x}) \to \delta \phi^i(t,\bm{x})\,;
\label{time_infdiffs}
\\
t&\to t\,, \qquad x^i \to x^i + \xi^i(t, \bm{x})\,, \qquad \delta \phi^i(t, \bm{x}) \to \delta \phi^i(t, \bm{x}) - \xi^i(t, \bm{x})\,,
\label{space_infdiffs}
\end{align}
where $\xi^0$ and $\xi^i$ are the parameters of the infinitesimal spacetime diffs. On the other hand, the infinitesimal internal diffs with the parameter~$\zeta^i(\phi)\simeq \zeta^i(\bm{x})$ read
\begin{align}
t&\to t\,, \qquad x^i \to x^i \,, \qquad \delta \phi^i(t, \bm{x}) \to \delta \phi^i(t, \bm{x})  + \zeta^i(\bm{x})
\,,
\label{internal_infdiffs}
\end{align}
where the condition to preserve the volume is given by $\partial_i \zeta^i=0$ with $\partial_i=\partial/\partial x^i$. One can notice that the fluctuations are invariant under the time-independent spatial diffs combined with the internal diffs by setting $\xi^i=\zeta^i(\bm{x})$:
\begin{align}
t\to t \,, \qquad x^i \to x^i + \zeta^i(\bm{x})\,, \qquad \delta\phi^i(t,\bm{x}) \to \delta \phi^i(t,\bm{x})
\,,
\end{align}
which we denote by $\diff$. We emphasize that the fluids do not respect the full $\diff$ symmetry and only the volume-preserving part of $\diff$, i.e., $\partial_i\zeta^i=0$, which we call $\Vdiff$, is the actual symmetry of the fluids.

The four-velocity takes $\langle u^{\mu} \rangle \propto \delta^{\mu}_0$ on the background and the flow (three-)velocity is given by $u^i \propto \partial_0 \delta \phi^i$ with $\partial_0=\partial/\partial t$. In the Eulerian description, the dynamics of the continuum is described by the evolution of $\delta \phi^i$ in terms of given spacetime coordinates. The EFT presented in \cite{Dubovsky:2005xd,Endlich:2010hf,Dubovsky:2011sj,Endlich:2012pz} is based on the Eulerian description. On the other hand, we shall adopt the Lagrangian description to construct the EFT of gravitating continuous materials which is more useful to extend the EFT into higher-orders. From the EFT perspective, adopting the Lagrangian description (the comoving gauge condition~$u^{\mu}\propto \delta^{\mu}_0$) is equivalent to choosing the unitary gauge~$\delta \phi^i=0$ by using the freedom of the spatial diffs in which the dynamical degrees of freedom of the continuum are eaten by the spacetime metric~$g_{\mu\nu}$. In the following, we use the terminology unitary gauge when we emphasize $\delta \phi^i=0$ while we call it the comoving gauge when emphasizing $u^{\mu}\propto \delta^{\mu}_0$, but these gauge conditions are equivalent in the present context. While the internal diffs do not keep the unitary/comoving gauge condition, $\diff$ preserves the conditions~$\delta \phi^i=0$ and $u^{\mu}\propto \delta^{\mu}_0$. The solid, the fluid, and the aether are thus distinguished by the (non-)invariance under $\diff$ in the unitary/comoving gauge. All hydrodynamical quantities are also read by the spacetime metric as we will show in the next subsection.

%%%%%%%%%%%%%%%%%%%%%%%%%%%%%%%%%%%%%%%%%%%%%%%%%%%%%%%%%%%
%%%%%%%%%%%%%%%%%%%%%%%%%%%%%%%%%%%%%%%%%%%%%%%%%%%%%%%%%%%
%%%%%%%%%%%%%%%%%%%%%%%%%%%%%%%%%%%%%%%%%%%%%%%%%%%%%%%%%%%
%%%%%%%%%%%%%%%%%%%%%%%%%%%%%%%%%%%%%%%%%%%%%%%%%%%%%%%%%%%

\subsection{The thread-based decomposition of spacetime}
\label{sec:hydro_decomposition}
Since the space and time are distinguished in the broken phase as seen from \eqref{time_infdiffs} and \eqref{space_infdiffs}, it is convenient to perform the $1+3$ decomposition of the spacetime. We often employ the ADM decomposition of the spacetime in which the spacetime is foliated with a family of constant-time hypersurfaces. However, the ADM decomposition is not useful in the present context, where the flow~$u^{\mu}$, rather than a constant-time hypersurface, plays an essential role. In fact, the four-velocity is invariant under the internal diffs so $u^{\mu}$ can be used as a building block of the EFT of any type of continuum. We thus adopt an alternative formulation of the $1+3$ decomposition with respect to the vector~$u^{\mu}$ which we have called the thread-based decomposition. See e.g.~\cite{Perjes:1992np,Boersma:1994pc,ellis2012relativistic,Roy:2014lda,Crossley:2015evo,Aoki:2021wew} for related studies and Appendix~\ref{app:1+3} for a few comments. The purpose of this subsection is to summarize the thread-based decomposition in a self-contained way. The essential results and the important quantities have been already explained in Sec.~\ref{introduction} and Table~\ref{table:notation}.

We first summarize hydrodynamical quantities in a spacetime-covariant way. The projection tensor orthogonal to $u^{\mu}$ is defined by
\begin{align}
h^{\mu\nu}\equiv g^{\mu\nu}+u^{\mu}u^{\nu}
\,.
\end{align}
The kinematical quantities of the flow (congruence), namely the expansion tensor, the vorticity tensor and the acceleration vector, are then defined by
\begin{align}
\expansion_{\mu\nu} &\equiv h_{(\mu|}{}^{\alpha} \nabla_{\alpha} u_{|\nu)} \,,  
\label{expansion} \\
\omega_{\mu\nu} &\equiv h_{[\mu|}{}^{\alpha} \nabla_{\alpha} u_{|\nu]}
\,, \label{vorticity} \\
a_{\mu} & \equiv u^{\alpha}\nabla_{\alpha} u_{\mu}
\,, \label{acceleration}
\end{align}
which completely describe the evolution of $u^{\mu}$:
\begin{align}
\nabla_{\mu}u_{\nu} = \expansion_{\mu\nu}+\omega_{\mu\nu}-u_{\mu}a_{\nu}
\,.
\end{align}
The expansion scalar and the shear tensor are the trace part and the traceless part of $\expansion_{\mu\nu}$,
\begin{align}
\expansion \equiv  \expansion^{\mu}{}_{\mu}\,, \qquad \sigma_{\mu\nu} \equiv \expansion_{\mu\nu}-\frac{1}{3}\expansion h_{\mu\nu}
\,.
\end{align}
The kinematical quantities are spatial quantities in the sense that they are orthogonal to $u^{\mu}$, e.g.~$\expansion_{\mu\nu}u^{\mu}=0$, so their indices can be raised and lowered by either $g^{\mu\nu}$ or $h^{\mu\nu}$. 

The dynamics of the system under consideration is governed by the action
\begin{align}
S[g_{\mu\nu},\phi^i]=S_{\rm EH}[g_{\mu\nu}] +S_{\rm m}[g_{\mu\nu},\phi^i]\,, \qquad S_{\rm EH}[g_{\mu\nu}]=\frac{\Mpl^2}{2}\int \D^4 x \sqrt{-g} R[g_{\mu\nu}]\,,
\end{align}
where $R[g_{\mu\nu}]$ is the spacetime Ricci scalar and $S_{\rm m}[g_{\mu\nu},\phi^i]$ is the action for the material.\footnote{For simplicity, we do not include higher-curvature corrections to the Einstein-Hilbert action.} We find the energy-momentum tensor by the variation of the matter action with respect to the metric,
\begin{align}
    T^{\mu\nu}\equiv \frac{2}{\sqrt{-g}}\frac{\delta S_{\rm m}}{\delta g_{\mu\nu}}
    \,.
\end{align}
The relativistic energy density~$\rho$, the relativistic energy flux (or the momentum density)~$q^{\mu}$ and the relativistic stress tensor~$\tau^{\mu\nu}$ are defined respectively by
\begin{align}
\rho \equiv T^{\alpha\beta}u_{\alpha}u_{\beta}\,, \qquad q^{\mu}\equiv - h^{\mu}{}_{\alpha}u_{\beta}T^{\alpha\beta} \,, \qquad
\tau^{\mu\nu}\equiv h^{\mu}{}_{\alpha} h^{\nu}{}_{\beta} T^{\alpha\beta} \,.
\end{align}
The trace part and the traceless part of $\tau^{\mu\nu}$ correspond to the isotropic stress~$p$ and the anisotropic stress~$\pi^{\mu\nu}$:
\begin{align}
    p \equiv \frac{1}{3}h_{\alpha\beta}\tau^{\alpha\beta}\,, \qquad \pi^{\mu\nu} \equiv \tau^{\mu\nu} - \frac{1}{3} (h_{\alpha\beta}\tau^{\alpha\beta}) h^{\mu\nu}
    \,.
\end{align}
The energy-momentum tensor is then expressed by
\begin{align}
    T^{\mu\nu} &=\rho u^{\mu}u^{\nu} + 2u^{(\mu}q^{\nu)} + \tau^{\mu\nu} \nn
    &=\rho u^{\mu}u^{\nu} + 2u^{(\mu}q^{\nu)} + p h^{\mu\nu} + \pi^{\mu\nu}
    \,.
    \label{Tmunu_dec}
\end{align}

We will mainly use the component expressions of tensors, i.e., tensors with indices such as $u^{\mu}$ and $T^{\mu\nu}$. On the other hand, we will choose the unitary gauge~$\delta \phi^i=0$ which is essentially a change of the basis from the Eulerian frame to the Lagrangian (comoving) frame. Since components of a tensor are not invariant under basis transformations, it is sometimes useful to deal with the tensor itself by writing the basis explicitly, e.g., $u^{\mu}\partial_{\mu}$ where $\partial_{\mu}$ is the spacetime holonomic (coordinate) basis that is the standard basis in the Eulerian frame. We will introduce another basis with the indices~$i,j,k,\cdots$ which denotes the basis in the Lagrangian frame.

We now adopt the unitary gauge~$\delta \phi^i=0$ in which the four-velocity is given by
\begin{align}
u^{\mu}&=\frac{1}{\sqrt{-g_{00}} } \delta^{\mu}_0\,,
\end{align}
where $g_{00}$ is the time-time component of $g_{\mu\nu}$. As we mentioned above, the unitary gauge is thus equivalent to the comoving gauge with respect to the flow (the Lagrangian description of the flow). We recall that the time diffs and $\diff$, which we call the thread-preserving diffs, preserve the unitary/comoving gauge condition:
\begin{align}
t\to t'(t,\bm{x})\,, \qquad x^i \to x'{}^i(\bm{x})\,, \qquad 
u^{\mu}\partial_{\mu}=\frac{1}{\sqrt{-g_{00}} } \partial_0 \to \frac{1}{\sqrt{-g_{00}} } \partial_0
\,.
\end{align}
We parametrize the difference between a future directed vector~$u_{\mu}$ and the vector normal to the constant-time hypersurface~$\delta^0_{\mu}=\partial_{\mu}t$ by using a positive function~$U=\sqrt{-g_{00}}$ and $U_i$:
\begin{align}
u_{\mu}=-U ( \delta^0_{\mu} - U_i \delta^i_{\mu})
\,.
\end{align}
This parametrization is particularly useful because it will turn out that the variables~$U$ and $U_i$ are conjugate to the energy density and the flux [see \eqref{energy} and \eqref{flux}].
Then, the $1+3$ form of the metric is given by
\begin{align}
g_{\mu\nu} &=
\begin{pmatrix}
-U^2 & U^2 U_j \\
U^2 U_i & h_{ij}-U^2 U_i U_j
\end{pmatrix}
\,, \label{glowu} \\
g^{\mu\nu} &=
\begin{pmatrix}
-\frac{1}{U^2}+U_k U^k & U^j \\
U^i & h^{ij}
\end{pmatrix}
\,, \label{gupu}
\end{align}
where $h_{ij}$ is the inverse matrix of $h^{ij}=\delta^i_{\mu}\delta^j_{\nu}h^{\mu\nu}$ and $U^i=h^{ij}U_j$.
The projection tensor with one upper index and one lower index is given by
\begin{align}
h^{\mu}{}_{\nu}=\left(\delta^{\mu}_i + U_i \delta^{\mu}_0 \right) \delta^i_{\nu}
\,.
\end{align}
The geometrical meaning of our $1+3$ decomposition may be clarified by writing the inverse of the spacetime metric~$g^{\mu\nu}$ as
\begin{align}
g^{\mu\nu}\partial_{\mu} \otimes \partial_\nu &= -\frac{1}{U^2} \partial_0 \otimes \partial_0 + h^{ij}\left( \partial_i + U_i \partial_0 \right) \otimes \left( \partial_j +U_j \partial_0 \right) 
\nn
&= -\dertau \otimes \dertau + h^{ij} \spatiald_i \otimes \spatiald_j
\,,
\end{align}
where the symbol~$\otimes$ denotes the tensor product and 
\begin{align}
\dertau \equiv u^{\mu}\partial_{\mu} = \frac{1}{U}\partial_0 \,, \qquad \spatiald_i \equiv h^{\mu}{}_i \partial_{\mu}=\partial_i + U_i \partial_0
\,.
\end{align}
Hence, the spacetime metric is decomposed into the direction parallel to the flow~$\dertau$ and the orthogonal directions~$\spatiald_i$ with (the inverse of) the ortho-spatial metric~$h^{ij}$. Physically, the differential operator~$\dertau$ represents the material derivative (i.e., the derivative with respect to the proper time of the comoving observer), while $\spatiald_i$ is the partial derivative with respect to the orthogonal directions~$\delta^{\perp} x^{\mu} = h^{\mu}{}_m \delta x^m=\left( U_n\delta x^n, \delta x^i \right)$. The transformation properties of $\dertau, \spatiald_i$ and $h^{ij}$ under the thread-preserving diffs are given by
\begin{align}
\dertau \to \dertau \,, \qquad \spatiald_i \to \frac{\partial x{}^j}{\partial x'{}^i} \spatiald_j
\,, \qquad h^{ij} \to \frac{\partial x'{}^i}{\partial x{}^k}\frac{\partial x'{}^j}{\partial x{}^l}h^{kl}
\,,
\end{align}
while $U$ and $U_i$ are transformed as
\begin{align}
U \to  \left( \frac{\partial t'}{\partial t} \right)^{-1} U \,, \qquad U_i \to \left( \frac{\partial x^j}{\partial x'{}^i}\right) \left[ \frac{\partial t'}{\partial t} U_j + \frac{\partial t'}{\partial x^j} \right]\,.
\label{U_transformations}
\end{align}

Any spacetime tensors can be decomposed into components parallel and orthogonal to $u^{\mu}$; for instance, a spacetime vector is decomposed as $V^{\mu}=V_{\parallel} u^\mu + \spatialV^{\nu} h^{\mu}{}_{\nu}$ with $V_{\parallel}\equiv -V^{\mu}u_{\mu}$ and $\spatialV^{\mu}\equiv h^{\mu}{}_{\nu}V^{\nu}$ [another example was shown in \eqref{Tmunu_dec}]. In particular, the comoving gauge (the Lagrangian frame) gives
\begin{align}
V^{\mu}\partial_{\mu} &= V_{\parallel} u^{\mu}\partial_{\mu} + \spatialV^{\nu} h^{\mu}{}_{\nu} \partial_{\mu} 
\nn
&=V_{\parallel} \dertau + \spatialV^i \spatiald_i\,,
\label{decV}
\end{align}
which shows that $V_{\parallel}$ and $\spatialV^i$ transform respectively as a scalar and a vector under 
$\diff$ and are invariant under the time diffs. Here, we emphasize that $\spatialV^{\mu}$ are the components of the vector in term of the holonomic basis (the Eulerian frame) with the orthogonality condition~$\spatialV^{\mu}u_{\mu}=0$ whereas $\spatialV^i$ should be regarded as the components in terms of the non-holonomic basis~$\spatiald_i$ (the Lagrangian frame). 
The first equality of \eqref{decV} holds for any choice of $u^{\mu}$ and the second line is the expression in the comoving gauge. It would be worthwhile mentioning the relations
\begin{align}
\spatialV^i = h^i{}_{\mu}\spatialV^{\mu}= \delta^i_{\mu} \spatialV^{\mu}\,, \qquad \spatialV^{\mu}=h^{\mu}{}_i \spatialV^i
\,,
\label{Vrel}
\end{align}
where the equalities hold only in the comoving gauge~$u^{\mu}=U^{-1}\delta^{\mu}_0$. 
As for a covector~$W_{\mu}$, the temporal component~$W_{\mu}u^{\mu}$ behaves as a scalar under $\diff$ while we obtain an ortho-spatial covector with the dual basis~$\D x^i= \D \phi^i$ from the orthogonal projection~$\spatialU_{\mu}\equiv h^{\nu}{}_{\mu}W_{\nu}$:
\begin{align}
\spatialU_{\mu} \D x^{\mu} = \spatialU_i \D x^i
\,.
\end{align}
Here, we have
\begin{align}
\spatialU_{\mu}=h^i{}_{\mu}\spatialU_i = \delta^i_{\mu} \spatialU_i \,, \qquad \spatialU_i=h^{\mu}{}_i \spatialU_{\mu}
\,,
\label{Urel}
\end{align}
in the comoving gauge. We can use \eqref{Vrel} and \eqref{Urel} to relate tensors with the Greek indices and those with the Latin indices under the unitary/comoving gauge condition~$u^{\mu}=U^{-1}\delta^{\mu}_0~(\delta \phi^i=0)$. 
The indices~$i,j,k,\cdots$ are raised and lowered by $h^{ij}$ and $h_{ij}$ which is verified by computing, e.g.,
\begin{align}
\spatialV^i =  \delta^i_{\mu}\spatialV^{\mu} = \delta^i_{\mu}h^{\mu\nu}\spatialV_{\nu} = h^{\mu\nu}\delta^i_{\mu}\delta^j_{\nu} \spatialV_j = h^{ij}\spatialV_j
\,.
\end{align}
Recall that the kinematical quantities are ortho-spatial quantities, yielding
\begin{align}
\expansion_{\mu\nu} \D x^{\mu} \otimes \D x^{\nu} = \expansion_{ij} \D x^i \otimes \D x^j
\,,
\end{align}
and the same is equally true for the other kinematical quantities. In the thread-based decomposition, we shall deal with quantities with the ortho-spatial indices~$i,j,k,\cdots$ such as $h^{ij}$ and kinematical quantities~$\{\expansion_{ij}, \omega_{ij}, a_i, \sigma_{ij} \}$ which have to be understood as (ortho-spatial) tensors in terms of the basis~$\spatiald_i$ and its dual~$\D x^i$, rather than the usual holonomic basis~$\partial_i$. These quantities are invariant under the time diffs and are tensorial under $\diff$. On the other hand, quantities without indices such as $V_{\parallel}$ and $\expansion$ are scalars under both time diffs and $\diff$.

The non-holonomicity of the basis indeed has the physical meaning. The Lie brackets of the basis~$\dertau$ and $\spatiald_i$ are given by
\begin{align}
[\dertau, \dertau]=0
\,, \qquad
[\dertau, \spatiald_i ] = a_i \dertau
\,, \qquad
[\spatiald_i, \spatiald_j ] = 2\omega_{ij}\dertau
\,,
\label{algebra}
\end{align}
with the Jacobi identities
\begin{align}
\spatiald_{[i} \omega_{jk]} -\omega_{[ij}a_{k]} &=0
\,, \qquad
\dertau \omega_{ij} - \spatiald_{[i}a_{j]} =0
\,. \label{Jacobi}
\end{align}
The non-vanishing parts on the right-hand side of \eqref{algebra} respectively agree with the acceleration and the vorticity of the flow:
\begin{align}
    a_i &= U( \dertau U_i - \spatiald_i U^{-1}) = \partial_0 U_i + U_i \partial_0 \ln U  + \partial_i \ln U
    \,, \\
    \omega_{ij} &= U \spatiald_{[i} U_{j]} = U\left( \partial_{[i}U_{j]}+U_{[i} \partial_0 U_{j]} \right)
    \,.
\end{align}
The algebra of $\spatiald_i$ is not closed within itself when the flow has a vorticity~$\omega_{ij}\neq 0$ reflecting the absence of the three-dimensional hypersurface orthogonal to $u^{\mu}$ (the Frobenius theorem). The acceleration represents the non-commutativity between $\dertau$ and $\spatiald_i$ and it vanishes when the flow is geodesic.

We then discuss covariant differentiations under the thread-preserving diffs. The material derivative~$\dertau$ is already a covariant derivative with respect to the time direction. In particular, the expansion tensor~$\expansion_{ij}$ is given by
\begin{align}
\expansion_{ij}=\frac{1}{2} \dertau h_{ij}
\,.
\end{align}
On the other hand, $\spatiald_i$ is not a covariant operator similarly to the partial derivative. The connection compatible with the metric~$h_{ij}$ is given by
\begin{align}
\spatialgam^i{}_{jk}\equiv \frac{1}{2}h^{il}\left( \spatiald_{j} h_{kl} + \spatiald_k h_{jl} -\spatiald_l h_{jk} \right)
\,,
\label{def:spatialgam}
\end{align}
in the comoving gauge and then the ortho-spatial covariant derivatives are given by
\begin{align}
\spatialD_i \spatialV^j= \spatiald_i \spatialV^j + \spatialgam^j{}_{ki}\spatialV^k
\,, \qquad 
\spatialD_i \spatialU_j = \spatiald_i \spatialU_j - \spatialgam^k{}_{ji} \spatialU_k
\,.
\label{eqs:DV}
\end{align}
We introduce a notation~$\spatialD_X$ to express the covariant derivatives in a basis-free way where $X=X_{\parallel} u^{\mu}\partial_{\mu}+\spatialX^{\mu}\partial_{\mu}=X_{\parallel} \dertau + \spatialX^i \spatiald_i$, with $u_{\mu}\spatialX^{\mu}=0$, is a spacetime vector field. The relations between the Eulerian frame and the Lagrangian frame are
\begin{align}
\spatialD_{X} \spatialS &= X_{\parallel} \pounds_{u} \spatialS + \spatialX^{\mu}\nabla_{\mu}\spatialS = X_{\parallel} \dertau \spatialS + \spatialX^i \spatialD_i \spatialS \,, 
\label{DXS} \\
\spatialD_{X} \spatialV &=(X_{\parallel} \pounds_{u} \spatialV^{\alpha}+\spatialX^{\mu}\nabla_{\mu}\spatialV^{\alpha}) h^{\beta}{}_{\alpha} \partial_{\beta} = (X_{\parallel} \dertau \spatialV^i + \spatialX^j \spatialD_j \spatialV^i) \spatiald_i 
\,, \label{DXV} \\
\spatialD_{X} \spatialU &=  (X_{\parallel}\pounds_{u} \spatialU_{\alpha}+ \spatialX^{\mu}\nabla_{\mu}\spatialU_{\alpha}) h^{\alpha}{}_{\beta} \D x^{\beta} = (X_{\parallel} \dertau \spatialU_i +\spatialX^j \spatialD_j \spatialU_i) \D x^i
\,, \label{DXU}
\end{align}
for a scalar~$\spatialS$, an ortho-spatial vector~$\spatialV=\spatialV^{\mu}\partial_{\mu}=\spatialV^i \spatiald_i$ and an ortho-spatial covector~$\spatialU=\spatialU_{\mu}\D x^{\mu}=\spatialU_i \D x^i$. Here, $\nabla_{\mu}$ is the spacetime covariant derivative and $\pounds_{u}$ is the Lie derivative with respect to $u^{\mu}$, respectively. We also note that the spacetime Christoffel symbols~$\Gamma^{\mu}_{\nu\rho}$ and the ortho-spatial connection~$\spatialgam^i{}_{jk}$ are related by
\begin{align}
\spatialgam^i{}_{jk} = h^{i}{}_{\mu} h^{\alpha}{}_{j} h^{\beta}{}_{k} \Gamma^{\mu}_{\alpha\beta}
\,,
\end{align}
provided $u^{\mu}=U^{-1}\delta^{\mu}_0$.

We introduce the ortho-spatial curvature associated with $h_{ij}$:
\begin{align}
\spatialR^i{}_{jkl}\equiv 2\left( \spatiald_{[k|}\spatialgam^i{}_{j|l]}+ \spatialgam^i{}_{n[k|}\spatialgam^n{}_{j|l]} \right)
\,.
\end{align}
The corresponding basis-free definition may be given by
\begin{align}
\spatialR(\spatialX, \spatialY) \spatialV \equiv  \spatialD_{\spatialX} \spatialD_{\spatialY} \spatialV - \spatialD_{\spatialY} \spatialD_{\spatialX} \spatialV - \spatialD_{[\spatialX, \spatialY]} \spatialV 
\,,
\end{align}
where $\spatialX=\spatialX^i \spatiald_i$ and $\spatialY=\spatialY^i \spatiald_i$ are ortho-spatial vectors. In the comoving gauge, we find
\begin{align}
\spatialR(\spatialX, \spatialY) \spatialV &= 2\spatialX^k \spatialY^l \left( \spatialD_{[k} \spatialD_{l]} \spatialV^i - \spatiald_{[k} \spatiald_{l]} \spatialV^i  \right) \spatiald_i
\nn
&=\spatialX^k \spatialY^l \spatialR^i{}_{jkl} \spatialV^j \spatiald_i
\,,
\end{align}
meaning that the curvature is not the commutator of $\spatialD_i$. The presence of $\spatiald_{[k} \spatiald_{l]} $ is easily understood by the non-holonomicity of the basis. The ortho-spatial curvature satisfies the following symmetric properties
\begin{align}
\spatialR_{(ij) kl} = 2\expansion_{ij}\omega_{kl}
\,, \qquad
\spatialR_{ij(kl)} =0 \,, \qquad
\spatialR_{i[jkl]} =0
\,.
\label{curvature_symmetry}
\end{align}
The existence of the symmetric part of $\spatialR_{ijkl}$ in the first two indices implies a non-zero ``non-metricity.'' The appearance of the ``non-metricity'' can be understood by the fact that the temporal part of $\spatialD_X$ is not compatible with the metric, $\dertau h_{ij}\neq 0$. The ortho-spatial Ricci tensor and scalar are defined by
\begin{align}
\spatialR_{ij} \equiv \spatialR^{k}{}_{ikj}\,, \qquad
\spatialR \equiv \spatialR^{i}{}_{i}
\,.
\label{spatialRicci}
\end{align}
We note that the ortho-spatial Ricci tensor is not symmetric, 
\begin{align}
\spatialR_{[ij]}=\expansion \, \omega_{ij}
\,.
\end{align}
We can also define the so-called co-Ricci tensor~$\spatialR^{i}{}_{kjl}h^{kl}$ and the homothetic curvature~$\spatialR^k{}_{kij}$, but they are represented by the ortho-spatial Ricci tensor and the kinematical quantities as clearly seen from \eqref{curvature_symmetry} and \eqref{spatialRicci}.

All the components of the four-dimensional curvature can be written in terms of the kinematical quantities and $\spatialR^i{}_{jkl}$ according to the Gauss, Codazzi, and Ricci equations
\begin{align}
    h^{\alpha}{}_{i} h^{\beta}{}_{j} h^{\gamma}{}_{k} h^{\delta}{}_{l} R_{\alpha\beta\gamma\delta} &=
    \spatialR_{ijkl} + 2 \spatialB_{[k|i} \spatialB_{|l]j}-2\spatialB_{ij}\omega_{kl}
    \,, \label{eqn:Gaussequation} \\
h^{\alpha}{}_{i} u^{\beta} h^{\gamma}{}_{k} h^{\delta}{}_{l} R_{\alpha\beta\gamma \delta } &=
2\spatialD_{[k} \spatialB_{l]i}-2a_{i} \omega_{kl}
\,,
\label{eqn:Codazziequation} \\
h^{\alpha}{}_{i} u^{\beta} h^{\gamma}{}_{k}u^{\delta} R_{\alpha\beta\gamma \delta } &= - \dertau \spatialB_{ki} + \spatialB_{i l} \spatialB_{k}{}^{l} + \spatialD_{k} a_{i} + a_{i} a_{k}
\,,
\label{eqn:Ricciequation}
\end{align}
where $\spatialB_{ij}\equiv \expansion_{ij}+\omega_{ij}$. In particular, the Einstein tensor~$G_{\mu\nu}$ and the Ricci scalar~$R$ are written as
\begin{align}
u^{\alpha}u^{\beta} G_{\alpha\beta}&= \frac{1}{2}\left( \spatialR-\expansion_{ij}\expansion^{ij}+\expansion^2 + 3\omega_{ij}\omega^{ij} \right)
\,, 
\label{decG00} \\
h^{\alpha}{}_{i}u^{\beta}G_{\alpha\beta}&=  \spatialD^j \spatialB_{ij} -\spatialD_i \expansion  + 2 a^j \omega_{ij}
\,, 
\label{decG0i} \\
h^{\alpha}{}_i h^{\beta}{}_j G_{\alpha\beta} &= \dertau \expansion_{ij} - h_{ij} \dertau \expansion - 2 \spatialB^{k}{}_{(i}\spatialB_{j)k} + \expansion \expansion_{ij}+ \spatialR_{(ij)} -\spatialD_{(i}a_{j)} - a_i a_j
\nn
&\quad -\frac{1}{2}h_{ij}(\spatialB_{kl}\spatialB^{kl}+\expansion^2+\spatialR - 2\spatialD_k a^k - 2a_k a^k)\,,
\label{decGij}
\end{align}
and
\begin{align}
    R&=\spatialR + \expansion_{ij} \expansion^{ij}+\expansion^2  +\omega_{ij}\omega^{ij} +2\dertau \expansion -2 \spatialD_i a^i-2a^i a_i
    \nn
    &=\spatialR + \expansion_{ij} \expansion^{ij}-\expansion^2  +\omega_{ij}\omega^{ij} -2 \nabla^{\mu}(a_{\mu} -\expansion u_{\mu})
\,,
\label{Ricci_relation}
\end{align}
with $a_{\mu}=\delta^i_{\mu}a_i$.

Finally, we discuss the Einstein equation and the energy-momentum conservation based on the thread-based decomposition. One can notice that the variables~$U,U_i$ and $h_{ij}$ are conjugate to the energy, the flux and the stress tensor:
\begin{align}
    \rho &= -\frac{U}{\sqrt{-g}} \frac{\delta S_{\rm m}}{\delta U} = -\frac{1}{\sqrt{h}} \frac{\delta S_{\rm m}}{\delta U}
    \,, \label{energy} \\
    q^i &= \frac{1}{U\sqrt{-g}} \frac{\delta S_{\rm m}}{\delta U_i} = \frac{1}{U^2 \sqrt{h}} \frac{\delta S_{\rm m}}{\delta U_i}
    \,, \label{flux} \\
    \tau^{ij} &= \frac{2}{\sqrt{-g}} \frac{\delta S_{\rm m}}{\delta h_{ij}} = \frac{2}{U\sqrt{h}} \frac{\delta S_{\rm m}}{\delta h_{ij}}
    \,,
\end{align}
where $h={\rm det}h_{ij}$.
Therefore, using the decomposition of the Einstein tensor~\eqref{decG00}--\eqref{decGij}, the Einstein equations are straightforwardly obtained once we specify a concrete form for the action~$S_{\rm m}=S_{\rm m}[U,U_i,h_{ij}]$ in the unitary gauge. 
As we will discuss later, it is useful to further decompose $h_{ij}$ into the determinant~$h={\rm det}h_{ij}$ and the unimodular metric~$\hat{h}_{ij}$:
\begin{align}
    h^{ij}=h^{-1/3} \hat{h}^{ij}\,, \qquad h_{ij}=h^{1/3} \hat{h}_{ij}
    \,,
\end{align}
where $\hat{h}^{ij}$ is the inverse of $\hat{h}_{ij}$ and ${\rm det}\hat{h}_{ij}={\rm det}\hat{h}^{ij}=1$. Treating $h$ and $\hat{h}_{ij}$ as independent variables, the variations of the action yield
\begin{align}
    p &= \frac{2h}{\sqrt{-g}}  \frac{\delta S_{\rm m}}{\delta h} =\frac{2h^{1/2}}{U} \frac{\delta S_{\rm m}}{\delta h}
    \,, \\
    \pi^{ij} &= \frac{2h^{-1/3}}{\sqrt{-g}}\left[ \delta^i_k \delta^j_l - \frac{1}{3}\hat{h}^{ij}\hat{h}_{kl}\right] \frac{\delta S_{\rm m}}{\delta \hat{h}_{kl}}
    \nn
    &=\frac{2}{U h^{5/6}}\left[ \delta^i_k \delta^j_l - \frac{1}{3}\hat{h}^{ij}\hat{h}_{kl}\right] \frac{\delta S_{\rm m}}{\delta \hat{h}_{kl}}
    \,,
\end{align}
so they are conjugate to the isotropic stress and the anisotropic stress, respectively. 
The hydrodynamical equations are read by the conservation law~$\nabla_{\mu}T^{\mu\nu}=0$:
\begin{align}
\mathcal{E}_{\parallel} &\equiv -(\nabla_{\mu}T^{\mu}{}_{\nu})u^{\nu}=\dertau \rho +\expansion (\rho+ p) +  \pi^{ij}\sigma_{ij}+\spatialD_i q^i + 2 a_i q^i =0\,, 
\label{energyconservation} \\
\mathcal{E}^{\perp}_i &\equiv (\nabla_{\mu}T^{\mu}{}_{\nu})h^{\nu}_i= (\rho+p)a_i + \spatialD_i p + h_{ij} \dertau q^j +2q^j \spatialB_{ji} + a^j \pi_{ij}+\spatialD^j \pi_{ji}=0
\,.
\label{momentumconservation}
\end{align}
As for the perfect fluid $\pi^{ij}=0=q^i$, the equations are reduced to the continuity equation and the Euler equation,
\begin{align}
\dertau \rho +\expansion(\rho + p)=0
\,, \qquad
(\rho+p)a_i+\spatialD_i p=0
\,.
\label{Eulereq}
\end{align}

The square root of the determinant~$h^{1/2}=({\rm det} h_{ij})^{1/2}$ is the volume element of the Lagrangian frame. Therefore, the quantity~$n=h^{-1/2}$ is interpreted as the number density of the fluid particles. The current is then $J^{\mu}= n u^{\mu}$ of which the conservation~$\nabla_{\mu}J^{\mu}=0$ identically holds.

We assume that the background is spatially flat, $\langle h_{ij} \rangle = a^2(t) \delta_{ij}$, which leads to $\langle \hat{h}_{ij} \rangle=\delta_{ij}$. For later convenience, we also define the quantity
\begin{align}
\delta \hat{h}_{ij} \equiv \hat{h}_{ij}-\delta_{ij}
\,,
\end{align}
which represents the perturbation of the unimodular metric around the background.

\subsection{Building blocks in the Lagrangian frame/unitary gauge}
\label{sec:building_blocks}

\begin{table}[t]
  \caption{Symmetries and building blocks in the Lagrangian frame. The one written in the upper row has a higher symmetry than the lower one as easily seen from the form of parameters ($c^i$ and $\Omega^i{}_j$ are a constant vector and a constant anti-symmetric matrix, respectively). Therefore, the EFT of fluid can use the building blocks of the aether as well and the EFT of solid can use those of the fluid.}
  \label{table:blocks}
  \centering
  \begin{tabular}{cccc}
      \hline
      Type &  Symmetry & Parameters &Essential building blocks \\
      \hline \hline
      Aether & 
      \begin{tabular}{l} Thread-preserving diffs: \\
      Time diffs + $\diff$ \end{tabular} & $\xi^0(t,\bm{x}),~\xi^i(\bm{x})$ &
      \begin{tabular}{l} $h^{ij},~\expansion_{ij},~\omega_{ij},~ a_i,$ \\ $\spatialR_{ijkl}, ~\dertau, ~\spatialD_i$ 
      \end{tabular} \\
      Fluid & Time diffs + $\Vdiff$ & $\xi^0(t,\bm{x}),~\xi^i(\bm{x})~{\rm s.t.}~\partial_i \xi^i=0$ & + $h$ \\
      Solid & Time diffs + $ISO_{\rm diag} (3)$ & $\xi^0(t,\bm{x}),~\xi^i=c^i+\Omega^i{}_j x^j$ & + $\delta \hat{h}_{ij}$ \\
      \hline
  \end{tabular}
\end{table}

In the previous subsection, we have explained the description of the spacetime in the Lagrangian frame of continuum (i.e.~the unitary/comoving gauge) and the relations to the hydrodynamical quantities. Since the thread-based decomposition manifestly respects the thread-preserving diffs, the formulation can be applied to all types of the continuum. In particular, the aether is defined as a phase of continuum that is invariant under all the thread-preserving diffs in the unitary/comoving gauge. The ordinary materials are, on the other hand, less symmetric than $\diff$; accordingly, we have a more freedom to construct the Lagrangian. In this subsection, we organize the building blocks of the EFT. The summary of this subsection is shown in Table~\ref{table:blocks}.

Let us start with building blocks of the aether. The Lagrangian of the aether must be given by a scalar that is invariant under the thread-preserving diffs and that is composed of the basic variables~$U,U_i$ and $h_{ij}$ in the unitary/comoving gauge. The transformation rule~\eqref{U_transformations} shows that $U$ and $U_i$ are not invariant under the time diffs; thus, the building blocks without the derivatives (which are equivalently the blocks consisting of first-order derivatives of $\phi^i$ in the Eulerian frame) are $h_{ij}$ and its inverse~$h^{ij}$. One may immediately notice that the only invariant scalar at this order is a constant. Hence, a non-trivial Lagrangian of aether has to have derivatives of $U,U_i$ and $h_{ij}$ (or the second- or higher-order derivatives of $\phi^i$). As we have discussed, the tensors under the thread-preserving diffs are the kinematical quantities~$\expansion_{ij},\omega_{ij},a_i$ at the first-order in derivatives and the ortho-spatial curvature~$\spatialR^i{}_{jkl}$ at the second-order. Higher-order quantities are obtained by taking the covariant differentiations~$\dertau$ and $\spatialD_i$. 

The fluid respects the time diffs and $\Vdiff$. Hence, the metric determinant~$h={\rm det}h_{ij}$ and the unimodular metric~$\hat{h}_{ij}=h^{-1/3}h_{ij}$ can be used as building blocks of the EFT of fluid. Since $\hat{h}_{ij}$ and $h_{ij}$ are linearly dependent, we can choose either of them as an independent building block. The material derivative of $h$ is related to the expansion scalar~$\expansion = \frac{1}{2}\dertau \ln h$ while $\spatialD_i h = \spatiald_i h$ gives a new building block of the EFT of fluids. The differential operator~$\spatialD_i$ is covariant under the generic $\diff$ but we do not necessarily respect the diffs that change the volume. One may introduce another covariant derivative~$\hat{\spatialD}_i$ which is compatible with $\hat{h}_{ij}$ in order to respect $\Vdiff$ only. The connection is defined as
\begin{align}
{}^{(3)}\! \hat{\gamma}^i{}_{jk}\equiv \frac{1}{2}\hat{h}^{il}\left( \spatiald_{j} \hat{h}_{kl} + \spatiald_k \hat{h}_{jl} -\spatiald_l \hat{h}_{jk} \right)
\,.
\end{align}
Nonetheless, $h_{ij}$ and $\hat{h}_{ij}$ are related in a conformal way and thus $\hat{\spatialD}_i$ can be written by $\spatialD_i$ and $\spatialD_i \ln h$. Therefore, the essentially new blocks of the EFT of fluids compared to the EFT of aether are $h$ and its ortho-spatial derivatives.

In the case of solids, we do not impose the $\diff$ symmetry. Since we have assumed the invariance under the spatial translation and rotation (the global $ISO_{\rm diag}(3)$ symmetry \eqref{ISOdiag}), the indices~$i,j,k,\cdots$ have to be contracted by either $h_{ij}$ or $\delta_{ij}$. In the EFT action, operators will be organized in powers of perturbations around the background. Due to the presence of two ``metrics''~$h_{ij}$ and $\delta_{ij}$, there are two possible contractions for each pair of indices, say $h_{ij}\spatialV^i \spatialV^j$ and $\delta_{ij}\spatialV^i \spatialV^j$, both of which are of order of $\mathcal{O}(\spatialV^2)$. On the other hand, as we assume a spatially flat background geometry, we can use $\delta\hat{h}_{ij}=\hat{h}_{ij}-\delta_{ij}$ as a building block alternative to $\delta_{ij}$. In this case, the quantity~$\delta \hat{h}_{ij}\spatialV^i \spatialV^j$ is of order of $\mathcal{O}(\delta \hat{h} \spatialV^2)$ which is higher order than $\mathcal{O}(\spatialV^2)$ in terms of the perturbations around the spatially flat background. One may worry about another contraction~$\delta^{ij}\spatialV_i \spatialV_j=\delta^{ij} h_{ik} h_{jl} \spatialV^k \spatialV^l$. However, by using the identity
\begin{align}
\delta^{[i|l}\delta^{|j|m}\delta^{|k]n}  =  \hat{h}^{[i|l}\hat{h}^{|j|m}\hat{h}^{|k]n}\,,
\label{del_identity}
\end{align}
where the $i,j,k$ indices are anti-symmetrized, $\delta^{ij}$ can be written in terms of $\delta \hat{h}_{ij}$ and $\hat{h}^{ij}$ as follows: 
\begin{align}
\delta^{il}=3\delta^{[i|l}\delta^{|j|m}\delta^{|k]n}\delta_{jm}\delta_{kn} 
=  3\hat{h}^{[i|l}\hat{h}^{|j|m}\hat{h}^{|k]n} (\hat{h}_{jm}-\delta \hat{h}_{jm})(\hat{h}_{kn} -\delta \hat{h}_{kn})
\,.
\end{align}
The material derivative of $\delta \hat{h}_{ij}$ is related to the shear tensor through $\frac{1}{2}\dertau \delta \hat{h}_{ij} = h^{-1/3}\sigma_{ij}$, and thus it does not yield a new operator of the EFT. Whereas, the ortho-spatial derivatives
\begin{align}
\spatialD_i \delta \hat{h}_{jk}=\spatiald_i \delta \hat{h}_{jk} - \spatialgam^l{}_{ji}\delta \hat{h}_{lk} - \spatialgam^l{}_{ki}\delta \hat{h}_{jl}
\end{align}
and
\begin{align}
\hat{\spatialD}_i \delta \hat{h}_{jk}=\spatiald_i \delta \hat{h}_{jk} - {}^{(3)}\! \hat{\gamma}^l{}_{ji}\delta \hat{h}_{lk} - {}^{(3)}\! \hat{\gamma}^l{}_{ki}\delta \hat{h}_{jl}
\end{align}
give new operators. Note that $\spatialD_i \delta \hat{h}_{jk}$ and $\hat{\spatialD}_i \delta \hat{h}_{jk}$ are not independent as we mentioned above. Similarly to the case of fluids, we do not need to use $\spatialD_i$ because the $\diff$ invariance is broken. One can use $\spatiald_i \delta \hat{h}_{jk}=\spatiald_i \hat{h}_{jk}$ as a building block of the EFT alternative to $\hat{\spatialD}_i \delta \hat{h}_{jk}$ (or $\spatialD_i \delta \hat{h}_{jk}$). Indeed, $\hat{\spatialD}_i \delta \hat{h}_{jk}$ is expressed in terms of $\spatiald_i \delta \hat{h}_{jk}$ and the other building blocks, and vice versa (and the same is true for $\spatialD_i \delta \hat{h}_{jk}$). It is straightforward to write $\hat{\spatialD}_i \delta \hat{h}_{jk}$ by means of $\spatiald_i \delta \hat{h}_{jk}$. To show the converse, we first express the connection~${}^{(3)}\! \hat{\gamma}^i{}_{jk}$ in terms of $\hat{\spatialD}_i \delta \hat{h}_{jk}$ by solving the equation
\begin{align}
\hat{\spatialD}_i \delta \hat{h}_{jk} = -\hat{\spatialD}_i \delta_{jk} = 2\delta_{l(j} {}^{(3)}\! \hat{\gamma}^l{}_{k)i}
\,.
\end{align}
We obtain
\begin{align}
{}^{(3)}\! \hat{\gamma}^i{}_{jk}=\frac{1}{2}\delta^{il}(\hat{\spatialD}_{j}\delta \hat{h}_{kl}+\hat{\spatialD}_k \delta \hat{h}_{jl}-\hat{\spatialD}_l \delta \hat{h}_{jk})\,,
\end{align}
and then find $\spatiald_i \delta \hat{h}_{jk}=2\hat{h}_{l(j}{}^{(3)}\! \hat{\gamma}^l{}_{k)i}$. All in all, the quantity~$\delta\hat{h}_{ij}=\hat{h}_{ij}-\delta_{ij}$ and its ortho-spatial derivatives are the new building blocks of the EFT of solids.

\subsection{Conserved currents}
\label{subsec:Noether}
We then discuss the consequence of Noether's theorem and find the corresponding conserved currents. Note that the current~$J^{\mu}=n u^{\mu}$ should be distinguished from the conserved currents discussed here. The conservation~$\nabla_{\mu}J^{\mu}=0$ is an off-shell identity arising from the fact that the number of the particles is fixed in the EFT (see \cite{Dubovsky:2005xd}). The conserved quantities in the perfect fluids and their consequences were examined in \cite{Dubovsky:2005xd}. The results can be easily extended to all orders of derivative expansions as well as the case of aether as we will see below.

In the unitary/comoving gauge, the matter action is given by a functional of the spacetime metric,
\begin{align}
S=S_{\rm EH}[g_{\mu\nu}]+S_{\rm m}[g_{\mu\nu}]
\,.
\end{align}
We consider the infinitesimal transformations
\begin{align}
x^{\mu} \to x^{\mu}+\xi^{\mu}\,, \qquad \xi^{\mu}=\xi_{\parallel} u^{\mu} + \xi_{\perp}^i h_i^{\mu}= \frac{1}{U}\xi_{\parallel} \delta^{\mu}_0 + \xi_{\perp}^i \delta^{\mu}_i
\,.
\label{Noether:transformation}
\end{align}
The gravity part $S_{\rm EH}$, namely the Einstein-Hilbert action, is invariant for any $\xi_{\parallel}=\xi_{\parallel}(t,\bm{x})$ and $\xi_{\perp}^i=\xi_{\perp}^i(t,\bm{x})$, which yields the contracted Bianchi identity~$\nabla_{\mu}G^{\mu}{}_{\nu}=0$ as usual. On the other hand, the matter action is invariant under only a part of the generic diffs: all types of continuum respect the time diffs~$\xi_{\parallel}=\xi_{\parallel}(t,\bm{x})$ whereas the aether and the fluids further enjoy the invariance under the generators of the residual symmetries~$\xi_{\perp}^i=\zeta^i(\bm{x})$ and $\xi_{\perp}^i=\zeta^i_V(\bm{x})$ s.t.~$\partial_i \zeta_V^i=0$, respectively. 

Let us change $x^{\mu}$ according to \eqref{Noether:transformation} without perturbing $x^{\mu}$ at the boundary of the spacetime. Note that the generator~$\xi_{\parallel}(t,\bm{x})$ can be chosen so that it vanishes at the spatial boundary and the initial and final constant-time hypersurfaces while $\xi_{\perp}^i=\zeta^i(\bm{x})$ does not vanish at the initial and final constant-time hypersurfaces because $\xi_{\perp}^i=\zeta^i(\bm{x})$ is a function of spatial coordinates only. The infinitesimal change of the matter action is
\begin{align}
\delta_{\xi}S_{\rm m} &= \frac{1}{2}\int \D^4x \sqrt{-g}T^{\mu\nu}\delta_{\xi}g_{\mu\nu} + {\rm boundary~terms}
\nn
&=\int \D^4 x \sqrt{-g}\left( - \mathcal{E}_{\parallel} \xi_{\parallel} + \mathcal{E}^{\perp}_i \xi_{\perp}^i \right) - \left[ \int \D^3 {\bm x} \mathcal{J}_i[g_{\mu\nu}] \xi_{\perp}^i \right]^{t=t_f}_{t=t_i}
,
\label{Noether:Sm}
\end{align}
with $\delta_{\xi}g_{\mu\nu}=-\pounds_{\xi}g_{\mu\nu}=-2\nabla_{(\mu}\xi_{\nu)}$ and a functional~$\mathcal{J}_i$. The last term represents the boundary term at the initial time~$t=t_i$ and the final time~$t=t_f$. Here we recall $\mathcal{E}_{\parallel}=-(\nabla_{\mu}T^{\mu}{}_{\nu})u^{\nu}$ and $\mathcal{E}^{\perp}_i=(\nabla_{\mu}T^{\mu}{}_{\nu})h^{\nu}{}_i$. Since the variation~\eqref{Noether:Sm} identically vanishes for any $\xi_{\parallel}=\xi_{\parallel}(t,\bm{x})$ with $\xi_{\perp}^i=0$, we obtain the identity 
\begin{align}
\mathcal{E}_{\parallel} \equiv 0
\,.
\end{align}
Therefore, the energy conservation~\eqref{energyconservation} is trivially satisfied for any type of the continuum, as it should be because we only have three scalar fields~$\phi^i$ when we do not impose the unitary/comoving gauge conditions, meaning that we only have three independent equations of motion in addition to the Einstein equations.

The aether respects the generic $\diff$. 
For the infinitesimal transformation~$\xi^i=\zeta^i(\bm{x})$ associated with $\diff$, \eqref{Noether:Sm} reads
\begin{align}
\delta_{\zeta} S_{\rm m}= \int \D^3 \bm{x} \zeta^i(\bm{x}) \int \D t \sqrt{-g} \mathcal{E}^{\perp}_i - \left[ \int \D^3 {\bm x} \mathcal{J}_i \zeta^i \right]^{t=t_f}_{t=t_i}\equiv 0\,,
\end{align}
which concludes
\begin{align}
 \int \D t \sqrt{-g} \mathcal{E}^{\perp}_i \equiv \mathcal{J}_i|_{t=t_f} - \mathcal{J}_i|_{t=t_i}
\,,
\label{Noetherid1}
\end{align}
since $\zeta^i(\bm{x})$ is an arbitrary function with compact support on the ortho-spatial manifold.
Note that \eqref{Noetherid1} is an off-shell identity independently of the initial and final constant-time hypersurfaces, implying
\begin{align}
\mathcal{E}^{\perp}_i = \frac{1}{\sqrt{h}}\dertau \mathcal{J}_i 
\,.
\end{align}
Hence, the momentum conservation~\eqref{momentumconservation} admits the first integral and $\mathcal{J}_i=\mathcal{J}_i(\bm{x})$ are the conserved currents associated with $\diff$.

The action of the fluid is invariant under the volume-preserving part of $\diff$. The infinitesimal transformation of $\Vdiff$ is written as $\xi_{\perp}^i=\zeta^i_V(\bm{x})=\varepsilon^{ijk}\partial_j \zeta^V_k(\bm{x})$. 
Then, the variation under $\Vdiff$ yields
\begin{align}
\delta_{\zeta^i_V} S_{\rm m} 
=\int \D^3 \bm{x} \zeta^V_k(\bm{x}) \int \D t \varepsilon^{ijk}\partial_i \left(\sqrt{-g}\mathcal{E}^{\perp}_j \right) 
- \left[ \int \D^3 {\bm x} \varepsilon^{ijk} \partial_i\mathcal{J}_j[g_{\mu\nu}] \zeta^V_k(\bm{x}) \right]^{t=t_f}_{t=t_i}\equiv 0
\,.
\end{align}
We thus obtain conserved currents from the transverse part of the momentum conservation. As mentioned in \cite{Dubovsky:2005xd}, this result leads to the relativistic generalization of Kelvin's theorem. Let us consider a perfect fluid described by the Lagrangian~$\mathcal{L}_{\rm m}=\rho_* F_0(h)$, where $\rho_*$ is a constant with the dimension of the energy density and $F_0(h)$ is an arbitrary dimensionless function. The energy density and the pressure respectively read $\rho = - \rho_* F_0$ and $p=\rho_*(F_0+2hF_h)$ with $F_h=\partial F_0/\partial h$ and the boundary term leads to 
\begin{align}
\partial_{[i}\mathcal{J}_{j]} = \partial_{[i} \left( \sqrt{-g}T^0{}_{j]} \right) =  \partial_{[i} \left[ \sqrt{h} (\rho+p) u_{j]} \right] 
,
\end{align}
where $u_i=U U_i$ are the spatial components of the four-velocity. One can directly check $\partial_{[i}\mathcal{J}_{j]} $ is conserved by calculating the transverse part of the Euler equation:
\begin{align}
\frac{1}{U} \partial_{[i}\left( \sqrt{-g} \mathcal{E}^{\perp}_{j]} \right) = \dertau \partial_{[i} \left[ \sqrt{h} (\rho+p) u_{j]} \right] =0
\,.
\end{align}
Considering a two-surface~$\Sigma$ bounded by a closed curve~$C$, we find
\begin{align}
\dertau \Gamma =0 \,, \qquad \Gamma \equiv  \oint_C \sqrt{h} \frac{\rho+p}{\rho_*} u_i \D x^i = \int_{\Sigma} \partial_{[i} \left[ \sqrt{h} \frac{\rho+p}{\rho_*} u_{j]} \right] \D x^i \wedge \D x^j
\,,
\end{align}
where $\Gamma$ is the circulation. The circulation reduces to $\Gamma = \oint_C u_i \D x^i$ in the case of the non-relativistic fluid satisfying $F_0= -h^{-1/2}=-n$. $\Vdiff$ implies that the degrees of freedom associated with fluid's vortex are integrable. 

Finally, we have $\xi^i=c^i+\Omega^i{}_j x^j$ for solids as is shown in Table~\ref{table:blocks}. In that case, as $c^i$ and $\Omega^i{}_j$ are constant, we recover the usual conservation of linear momentum and angular momentum corresponding to the residual symmetry~$ISO_{\rm diag} (3)$.

\subsection{Transformation to the Eulerian frame}
\label{sec:Eulerian}
So far, we have adopted the comoving coordinates with respect to the continuum in which the degrees of freedom of the continuum are eaten by the spacetime metric (the Lagrangian description). We now derive a relationship between the quantities in the Lagrangian frame and those in an arbitrary coordinate system (the Eulerian description). This is essentially just a coordinate transformation or in the EFT perspective this is the St\"{u}ckelberg trick to restore the spacetime covariance of the action. The degrees of freedom of the continuum described by $\phi^i$ (or $\delta \phi^i$) can be regarded as the Nambu-Goldstone bosons associated with the broken spatial diffs.

Since most of the quantities in the Lagrangian frame are components of tensors in the basis~$(\spatiald_i, \D x^i)$, the transformation of a scalar quantity is easily achieved by replacing the Latin indices with the Greek indices, for instance,
\begin{align}
\expansion_{ij}\expansion^{ij}=h^{ik}h^{jl}\expansion_{ij}\expansion_{kl} \to \expansion_{\mu\nu}\expansion^{\mu\nu}=h^{\mu\rho}h^{\nu\sigma}\expansion_{\mu\nu}\expansion_{\rho\sigma}
\,,
\end{align}
where $\expansion_{\mu\nu}$ is defined by \eqref{expansion}. The derivatives~$\dertau$ and $\spatialD_i$ can be also replaced according to \eqref{DXS}--\eqref{DXU}. On the other hand, $h={\rm det}h_{ij}$ and $\delta \hat{h}_{ij}$ are not tensors with respect to the generic $\diff$ and thus the simple replacement~$i,j,k,\cdots \to \mu,\nu,\rho,\cdots$ does not work for them. 

As we usually do in continuum mechanics, it is useful to express the Lagrangian-frame quantities in terms of the Eulerian ones, namely $g_{\mu\nu}$ and $\phi^i$. Let us introduce subscripts $L$ and $E$ to emphasize whether a quantity is expressed in terms of the Lagrangian-frame variables or the Eulerian-frame variables. The quantities with $L$ are understood as functionals of $U_L(t,\bm{x}_L),U_{Li}(t,\bm{x}_L),h_{Lij}(t,\bm{x}_L)$ as summarized in Table~\ref{table:notation} while the quantities with $E$ are functionals of $g_{E\mu\nu}(t,\bm{x}_E)$ and $\phi^i_E(t,\bm{x}_E)$. For instance, the expansion tensor in the Lagrangian frame is denoted by $\expansion_{L}^{ij}=\expansion_{L}^{ij}[U_L,h_{Lij}]$ and its Eulerian expression is $\expansion_{E}^{ij}=\expansion_{E}^{ij}[g_{E\mu\nu},\phi^i_E]$. Let us argue how we can find the expression of $\expansion_{E}^{ij}$ from $\expansion_{L}^{ij}$.

We recall that $h^{ij}_E$ is originally defined as
\begin{align}
h_E^{ij}\equiv g_E^{\mu\nu} \partial_{E \mu}\phi^i_E \partial_{E \nu}\phi^j_E = h^{\mu\nu}_E \partial_{E \mu}\phi^i_E \partial_{E \nu}\phi^j_E\,,
\end{align}
where $h^{\mu\nu}_E=g^{\mu\nu}_E+u^{\mu}_E u^{\nu}_E$. Its inverse matrix is explicitly computed by
\begin{align}
h_{Eji}=\frac{1}{2{\rm det}h_E^{pq}}\varepsilon_{ikl}\varepsilon_{jmn}h_E^{km}h_E^{ln} \,.
\end{align}
We then introduce the quantities
\begin{align}
h_E^i{}_{\mu} &\equiv  \partial_{E \mu}\phi^i_E \,, \\
h_E^{\mu}{}_i &\equiv h_{Eij} \partial_{E \nu}\phi^j_E g^{\mu\nu}_E
\,,
\end{align}
which satisfy
\begin{align}
h_E^i{}_{\mu} h_E^{\mu}{}_j = \delta^i{}_j
\,, \qquad h_E^{\mu}{}_i h_E^i{}_{\nu} = h_E^{\mu}{}_{\nu}
\,,
\end{align}
and
\begin{align}
h_E^i{}_{\mu}u^{\mu}_E=0\,, \qquad h_E^{\mu}{}_i u_{E\mu}=0\,.
\end{align}
For tensors orthogonal to $u^{\mu}_E$, we can change their Greek indices into the Latin indices and vice versa by the use of $h_E^i{}_{\mu} $ and $h_E^{\mu}{}_i $.
For instance, 
\begin{align}
\spatialV &=\spatialV^{\mu}_E \partial_{E \mu} = \spatialV^i_E \spatiald_{Ei}
\,, \\
\spatialU &= \spatialU_{E\mu} \D x^{\mu}_E = \spatialU_{Ei} \D \phi^i_E
\,,
\end{align}
where
\begin{align}
\spatialV^i_E\equiv \spatialV^{\mu}_E h^i_E{}_{\mu}\,, \qquad 
\spatialU_{Ei}\equiv  \spatialU_{E\mu} h^{\mu}_E{}_{i}
\,,
\end{align}
and
\begin{align}
\spatiald_{Ei} &\equiv h_E^{\mu}{}_i \partial_{E\mu}\,, \qquad \D \phi^i_E \equiv h^i_E{}_{\mu} \D x^{\mu}_E
\,.
\end{align}
The material derivative in the Eulerian frame is given by
\begin{align}
\tilde{\partial}_{E 0}\equiv u^{\mu}_E \partial_{E\mu}= u^0_E \partial_0 + u^i_E \partial_{E i}
\,.
\end{align}
Note that the expressions here are applicable to any coordinate systems since we have not chosen any specific coordinates to define the quantities with $E$. These quantities are reduced to what we have defined in Sec.~\ref{sec:hydro_decomposition} by setting $\phi_E^i=x^i_L$. In this way, one can easily express all the quantities of the Lagrangian frame in terms of $g_{E\mu\nu}(t,\bm{x}_E)$ and $\phi^i_E(t,\bm{x}_E)$. The transformation to the Eulerian frame is then straightforward: one can simply replace the quantities of the Lagrangian frame with those with $E$,
\begin{align}
\expansion_{L}^{ij}  &\to \expansion_{E}^{ij}=h^{i}_E{}_{\mu} h^{j}_E{}_{\nu} \expansion_{E}^{\mu\nu} = -\frac{1}{2}\tilde{\partial}_{E 0} h_{E}^{ij}
\,, \nn
h_L&\to h_E = {\rm det}h_{Eij}
\,, \qquad
\delta \hat{h}_{L ij} \to \delta \hat{h}_{E ij}=h_E^{-1/3}h_{Eij}-\delta_{ij}
\,,
\end{align}
and so on. For simplicity of the notation, we suppress the subscripts~$E$ and $L$ in the following. Whether we are using the Lagrangian description or the Eulerian description will be clear from the context or the difference is irrelevant for discussions.

%%%%%%%%%%%%%%%%%%%%%%%%%%%%%%%%%%%%%%%%%%%%%%%%%%%%%%%%%%%%%%%%%%%%%%%%%%%%%%%%%%%%
%%%%%%%%%%%%%%%%%%%%%%%%%%%%%%%%%%%%%%%%%%%%%%%%%%%%%%%%%%%%%%%%%%%%%%%%%%%%%%%%%%%%
%%%%%%%%%%%%%%%%%%%%%%%%%%%%%%%%%%%%%%%%%%%%%%%%%%%%%%%%%%%%%%%%%%%%%%%%%%%%%%%%%%%%
%%%%%%%%%%%%%%%%%%%%%%%%%%%%%%%%%%%%%%%%%%%%%%%%%%%%%%%%%%%%%%%%%%%%%%%%%%%%%%%%%%%%

\section{Derivative expansion}
\label{sec:derivative}

Let us study a concrete Lagrangian by using the formalism developed in Sec.~\ref{sec:generalities}; please refer to the summary of the building blocks provided in Table~\ref{table:blocks}. The discussion in Sec.~\ref{subsec:leading} does not rely on whether the description of the continuum is Lagrangian or Eulerian, while we will use the Eulerian description in Sec.~\ref{subsec:decoupling} to take the decoupling limit of gravity.

\subsection{Leading operators}
\label{subsec:leading}
We are interested in low-energy/momentum physics where microscopic properties of a material are negligible. We employ the derivative expansion and organize the Lagrangian in terms of the number of derivatives. Note that the first-order derivative of the Eulerian-frame variable~$\phi^i$ is of order unity at the background, $\langle \partial_{\mu} \phi^i \rangle = \delta^i_{\mu}$, implying that we cannot perform a perturbative expansion in terms of the first-order derivative of $\phi^i$. Our assumption is 
\begin{align}
g_{\mu\nu}, \partial \phi^i = \mathcal{O}(1)\,, \qquad \partial^n g_{\mu\nu}, \partial^{n+1} \phi^i \ll \Lambda^n\,, \quad (n=1, 2, \cdots)\,,
\end{align}
where $\Lambda$ is the cutoff of the EFT which would represent the microscopic scale of the continuum. This power counting is more transparent in terms of the Lagrangian-frame variables:
\begin{align}
U,U_i,h_{ij} = \mathcal{O}(1)\,, \qquad \dertau, \spatiald_i \ll \Lambda
\,.
\end{align}
We thus count the number of the derivatives using the quantities in the Lagrangian frame (the building blocks of the EFT in the unitary/comoving gauge).

At the zeroth-order in derivative, the building block of the EFT is $h_{ij}$. Allowed operators of the EFT depend on the symmetry. In the case of the solid (the lowest symmetric case in our consideration), the most general Lagrangian at the zeroth-order in derivative is given by a function~$F=F(\tr h, \tr h^2, \tr h^3)$ in three spatial dimensions according to the Cayley-Hamilton theorem~\cite{Endlich:2012pz}, where
\begin{align}
\tr h \equiv h^{ij}\delta_{ij}
\,, \qquad
\tr h^2 \equiv h^{ij}h^{kl}\delta_{jk}\delta_{li}
\,, \qquad
\tr h^3 \equiv h^{ij}h^{kl}h^{mn}\delta_{jk}\delta_{lm}\delta_{ni}
\,.
\label{traceh}
\end{align}
It is more convenient to use other combinations as the arguments of the function~$F$ in order to compare the EFT of solid with that of fluid. Let us introduce the notations
\begin{align}
[\delta \hat{h}] \equiv \hat{h}^{ij}\delta \hat{h}_{ij}
\,, \qquad
[\delta \hat{h}^2]  \equiv \hat{h}^{ij}\hat{h}^{kl}\delta \hat{h}_{jk}\delta \hat{h}_{li}
\,, \qquad
[\delta \hat{h}^3]  \equiv \hat{h}^{ij}\hat{h}^{kl}\hat{h}^{mn}\delta \hat{h}_{jk}\delta \hat{h}_{lm}\delta \hat{h}_{ni}
\,.
\label{tracehath}
\end{align}
Straightforward calculations yield
\begin{align}
6h^{-1} &= (\tr h)^3 - 3 (\tr h) (\tr h^2)+ 2 (\tr h^3)
\,, \\
[\delta \hat{h}] &= 3 - h^{1/3} (\tr h)
\,, \\
[\delta \hat{h}^2] &=3 - 2 h^{1/3} (\tr h) + h^{2/3} (\tr h^2) 
\,, \\
[\delta \hat{h}^3] &=3 - 3 h^{1/3} (\tr h) + 3 h^{2/3} (\tr h^2) - h (\tr h^3)
\,,
\end{align}
and the relation
\begin{align}
[\delta \hat{h}] - \frac{1}{2} \left([\delta \hat{h}]^2-  [\delta \hat{h}^2] \right)+\frac{1}{6}\left( [\delta \hat{h}]^3  - 3 [\delta \hat{h}] [\delta \hat{h}^2] + 2 [\delta \hat{h}^3] \right)=0
\,.
\end{align}

Let us optimize the basis to parametrize the function~$F$. The paper~\cite{Endlich:2012pz} used $\tr h, \tr h^2/(\tr h)^2, $ and $\tr h^3/(\tr h)^3$ where the last two become constant at the background, $\tr h^2/(\tr h)^2 \to 1/3$ and $\tr h^3/(\tr h)^3 \to 1/9$. Instead, one can use the determinant~$h={\rm det}h_{ij}$ and two combinations of \eqref{tracehath} to specify the function~$F$. Choosing $h$ as one of the building blocks is useful since it is also building block of the EFT of fluids. The most useful combinations for the rest would be the elementary symmetric polynomials
\begin{align}
e_2 &\equiv  \frac{1}{2} \left( [\delta \hat{h}]^2-  [\delta \hat{h}^2] \right)
\,, \\
e_3 & \equiv \frac{1}{6}\left( [\delta \hat{h}]^3  - 3 [\delta \hat{h}] [\delta \hat{h}^2] + 2 [\delta \hat{h}^3] \right)
\,,
\end{align}
as one can easily express $\{ \tr h, \tr h^2/(\tr h)^2, \tr h^3/(\tr h)^3 \}$ in terms of $\{h,e_2,e_3\}$.
In our basis $\{h,e_2,e_3\}$, only $h$ has a non-vanishing value at the background while $e_2$ and $e_3$ start at the quadratic and cubic orders in perturbations around the background, respectively. Hence, the action can be easily organized to the power of perturbations as well as to respect the symmetries of the fluids and the solids.

Then, the Lagrangian up to the zeroth-order in derivative is 
\begin{align}\label{zeroth-O}
\mathcal{L}_{\rm m}{\big{|}}_{\text{zeroth-order}}
=\rho_* F(h,e_2,e_3) \,,
\end{align}
where $\rho_*$ denotes the scale of the energy density of the continuum. The fluid corresponds to $F=F(h)$ or the limit
\begin{align}
\frac{\partial^p}{\partial e_2^p} \frac{\partial^q}{\partial e_3^q} F(h,e_2,e_3) \to 0 
\quad \text{for any non-negative integers~$p$ and $q$ with $p+q\geq 1$}
\,.
\end{align}
The aether does not allow to have a non-trivial Lagrangian at the zeroth-order in derivative and the only possible term of the aether is $F=$ constant, which is the cosmological constant, at this order.

The first-order operators in the derivative expansion, which are relevant for the Lagrangian, are $\expansion$ and the action of $\dertau$ on the zeroth-order quantities like $h$, $e_2$, $e_3$. In the case of fluids and aether, the possible first-order operators are $\mbox{constant}\times\expansion$ and $f(h)\expansion$ respectively, which are total derivative. On the other hand, solids generically admit non-trivial first-order operators such as $e_2 \expansion$.
As $\expansion = \frac{1}{2}\dertau\ln{h}$, all possible first-order terms include one material derivative. Then, if we additionally impose {\it the time reflection symmetry}, all possible first-order terms are prohibited even in the case of solids.
%\footnote{Even if we do not impose the time reflection symmetry, the possible terms for the case of aether and fluids are $\mbox{constant}\times\expansion$ and $f(h)\expansion$ respectively, which are total derivative.}

The non-trivial second-order operators which contain two derivatives and contribute up to the quadratic order in the Lagrangian of the perturbations are given by (see Appendix \ref{app:HDO} for more details)
\begin{align}
\label{2nd-O-O}
&\expansion^2\,, \quad \sigma_{ij}\sigma^{ij}\,, \quad \omega_{ij}\omega^{ij}\,, \quad a_i a^i\,, \quad \spatialR\,,
\\
\label{2nd-O-O-f}
&\spatialD_i h \spatialD^i h  \,, \quad a^i \spatialD_i h 
\,,
\\
\label{2nd-O-O-s}
&h^{ij} \divh_i \divh_j\,, \quad
\divh_i \spatialD^i h \,, \quad
\divh_i a^i
\,,
\end{align}
where we have defined
\begin{align}\label{Dh-sv}
\divh_i \equiv h^{jk}\spatiald_j \delta \hat{h}_{ki}
\,.
\end{align}
For simplicity, we have assumed the spatial parity invariance.
The first-order operators~${\spatialD}_i h$ and $\divh_i$ characterize the longitudinal and transverse parts of $h_{ij}$ while the transverse-traceless part is encoded in $\spatialR$. Note that when we construct the EFT Lagrangian up to the quadratic order in perturbations for solids and fluids, the coefficients of the above operators are general functions of only $h$ as $e_2$ and $e_3$ start at the quadratic and cubic orders in perturbations respectively. The operators in \eqref{2nd-O-O}, \eqref{2nd-O-O-f}, and \eqref{2nd-O-O-s} are relevant for the solids, while only \eqref{2nd-O-O} and \eqref{2nd-O-O-f} are relevant for the fluids since $\delta \hat{h}_{ij}$ is not a building block for the fluids. For the aether, neither $h$ nor $\delta \hat{h}_{ij}$ are building blocks and therefore only the operators in \eqref{2nd-O-O} are relevant and their coefficients in the Lagrangian should be constant as well. Consequently, the operator~$\spatialR$ can be eliminated by redefining the Planck mass of the Einstein-Hilbert action and the coefficients of the first three operators in \eqref{2nd-O-O} [recall that the full action of the present system is $S=S_{\rm EH}+S_{\rm m}$ and see also Eq.~\eqref{Ricci_relation}].

Although operators with derivatives, like the second-order ones~\eqref{2nd-O-O}, \eqref{2nd-O-O-f}, \eqref{2nd-O-O-s}, are sub-leading in low-energy/momentum dynamics of the solids and fluids, they become leading operators in the case of aether due to the high symmetry of the system. Thus, in order to be able to study perturbations of the aether, we have to keep the relevant second-order operators~\eqref{2nd-O-O}. 
As a result, the leading-order Lagrangian including the operators characterizing the aether is given by
\begin{align}
\mathcal{L}_{\rm m} 
&= \rho_* \left[ F(h,e_2,e_3) + \mathcal{O}\left(\frac{\partial}{\Lambda}\right) \right]
\nn
&\quad +\rho_* \left[ \frac{1}{2\Lamae^2}\left( b_{\expansion^2} \expansion^2 + b_{\sigma^2}\sigma_{ij}\sigma^{ij} + b_{\omega^2} \omega_{ij}\omega^{ij} + b_{a^2} a_i a^i \right)+ \mathcal{O}\left(\frac{\partial^3}{\Lamae^3}\right)\right]
,
\label{leadingL}
\end{align}
where the operators in the first line do not respect the generic $\diff$ invariance while those in the second line are invariant under $\diff$. Here, we additionally introduce the scale~$\Lamae$ to denote the scale of the aether operators. The hierarchy~$\Lamae\ll \Lambda $ is technically natural since the operators of the aether do not renormalize the operators of the solids and fluids thanks to the differences in the underlying symmetries. Hence, assuming either approximate or exact $\diff$ symmetry, we may ignore derivative operators in the first line of \eqref{leadingL} even when the second line becomes important.

\begin{table}[t]
  \caption{Classification of continua.}
  \label{table:classification}
  \centering
  \begin{tabular}{lll}
      \hline
      Type & Conditions \\
      \hline \hline
      Solid & No conditions \\
      Fluid & $ \partial^p_{e_2} \partial^q_{e_3} F(h,e_2,e_3) \to 0$ \\
      Non-ideal aether & $ \partial_h^p \partial^q_{e_2} \partial^r_{e_3} F(h,e_2,e_3) \ll 1 $ \\
      Integrable aether & $ \partial_h^p \partial^q_{e_2} \partial^r_{e_3} F(h,e_2,e_3) \to 0$ and $b_{a^2} \to 0$ \\
      Ideal aether & $ \partial_h^p \partial^q_{e_2} \partial^r_{e_3} F(h,e_2,e_3) = 0$ and $b_{a^2} \neq 0$ 
      \\
      \hline
  \end{tabular}
\end{table}

We give a classification of the continua in Table~\ref{table:classification}. We use the terms ``ideal'' and ``non-ideal'' to distinguish whether the symmetry is exact or approximate. In this definition, the ideal fluid may have a flux and an anisotropic stress due to higher-derivative corrections. For instance, one may consider $\spatiald_i n \spatiald^i n$, which comes from the second-order operator~${\spatialD}_i h{\spatialD}^i h$ in \eqref{2nd-O-O-f}, as a higher-derivative correction of the ideal fluid. Since $n=h^{-1/2}$ is interpreted as the number density, this operator represents an interaction arising from a difference of the number density between adjacent fluid particles, yielding a flux~$q^i \propto \expansion \spatiald^i n$ and an anisotropic stress~$\pi^{ij} \propto \spatiald^i n \spatiald^j n -\frac{1}{3}h^{ij} (\spatiald n)^2$. We especially call the leading-order Lagrangian of the ideal fluid the perfect fluid. The Lagrangian of the perfect fluid is $\mathcal{L}_{\rm m}=\rho_* F_0(h)$, which has neither the flux nor the anisotropic stress.
Also, the aether has three possible phases depending on whether the symmetry is approximate or exact.
The non-ideal aether refers to a continuum with the approximate $\diff$ invariance and the integrable aether appears as its ideal limit under a scaling of the coupling constant~$b_{a^2}$ as we will discuss later. The ideal aether is the case with the exact $\diff$ invariance. In this classification, the so-called Einstein-aether theory is categorized into a special case of the ideal aether (see Appendix~\ref{app:Einsteinaether} for the relation between our aether and the Einstein-aether theory). As we will see in the next subsection, the Nambu-Goldstone bosons, which can be interpreted as the phonons of the continuum, show different properties depending on the symmetry of the continuum. In particular, the aethers are quite peculiar and their property strongly depends on whether the symmetry is approximate or exact.

Before studying perturbations around the background, we shall discuss the property of the background. The background is not only homogeneous and isotropic in space but also static in time when gravity is turned off. The energy density and the pressure of the background are then
\begin{align}
    \bar{\rho} \equiv \langle \rho \rangle =  -\rho_* \langle F \rangle
    \,, \qquad \bar{p} \equiv \langle p \rangle = \rho_* \langle F + 2h F_h \rangle 
    \,,
\end{align}
where $F_h=\partial F/\partial h$ and $\langle \cdots \rangle$ means that the quantity inside the angle brackets is evaluated at the background. We shall omit the angle brackets to simplify the notation if no confusion arises. The equation of state parameter of the background is 
\begin{align}
    w \equiv \frac{\bar{p}}{\bar{\rho}} = -1 +\frac{2\rho_* h F_h}{\bar{\rho}} = -1 - \frac{2h F_h}{F}
    \,.
\end{align}
Hence, the aethers correspond to $w\simeq -1$ and the background exactly behaves as a cosmological constant in the ideal case where $F_h=0$. When the system gravitates, the background may not be static and then the expansion scalar does not vanish. Then, the total energy density and the pressure of the continuum are corrected by higher-derivative terms such as $\expansion^2$. Nonetheless, these effects are negligible at the leading order since the regime of validity of the EFT is limited to $|\expansion| \ll \Lamae$.

\subsection{Decoupling limit}
\label{subsec:decoupling}

We study the dynamics of the phonons described by the leading-order Lagrangian~\eqref{leadingL} under the decoupling limit of gravity~$\Mpl \to \infty$. We use the Eulerian frame and consider perturbations around the Minkowski background:
\begin{align}
g_{\mu\nu}=\eta_{\mu\nu} + \frac{1}{\Mpl} \delta g_{\mu\nu} \,, \qquad \phi^i = x^i + \delta \phi^i
\,,
\end{align}
where $\eta_{\mu\nu}$ is the Minkowski metric. The metric fluctuations are decoupled from $\delta \phi^i$ in the decoupling limit. We leave the analysis of gravitational perturbations to Sec.~\ref{sec:gravitating_continuum}.
Note that most of the results of the quadratic action are already known even in the presence of gravity (cf.~\cite{Endlich:2012pz}) except for the case of aether. Nonetheless, one may notice that the practical calculations, especially in the presence of gravity, are simplified thanks to the reformulation of the EFT action in the unitary/comoving gauge.

The action~\eqref{leadingL} leads to
\begin{align}
\mathcal{L}_{\rm m} 
= \rho_* \Big[ & F + F_h (h-1) + \frac{1}{2}F_{hh} (h-1)^2 + F_{e_2} e_2
\nn
&+\frac{1}{2\Lamae^2}\left( b_{\expansion^2} \expansion^2 + b_{\sigma^2}\sigma_{ij}\sigma^{ij} + b_{\omega^2} \omega_{ij}\omega^{ij} + b_{a^2} a_i a^i \right) + \cdots \Big]
\,,
\end{align}
where it is understood that the coefficients are evaluated on the background ($h=1$, $e_1=0$, $e_2=0$), and the ellipsis denotes terms that are irrelevant for the quadratic action of $\delta \phi^i$.
In the perturbative analysis, we do not distinguish upper indices and lower indices, rather than raising and lowering the indices by the background spatial metric~$\delta_{ij}$ and $\delta^{ij}$, and adopt the notation that an index variable appearing twice in subscript implies the summation. We obtain
\begin{align}
h&=1-2\partial_i \delta \phi_i + \delta \dot{\phi}_i \delta \dot{\phi}_i + 2 (\partial_i \delta \phi_i)^2 +\partial_i \delta \phi_j \partial_j \delta \phi_i
+ \mathcal{O}(\delta \phi^3)
\,, \\
\delta \hat{h}_{ij}&=-2\partial_{(i} \delta \phi_{j)} + \frac{2}{3}\delta_{ij} \partial_k \delta \phi_k + \mathcal{O}(\delta \phi^2)
\,,
\end{align}
where a dot is the derivative with respect to $t$. The four-velocity is given by
\begin{align}
u^{\mu}=(1, -\delta \dot{\phi}_i) + \mathcal{O}(\delta \phi^2)
\,,
\end{align}
and then the kinematical quantities are
\begin{align}
\expansion_{ij} &=-\partial_{(i} \delta \dot{\phi}_{j)}  + \mathcal{O}(\delta \phi^2)
\,, \\
\omega_{ij} &=-\partial_{[i} \delta \dot{\phi}_{j]}  + \mathcal{O}(\delta \phi^2)
\,, \\
a_i &= - \delta \ddot{\phi}_i  + \mathcal{O}(\delta \phi^2)
\,.
\end{align}
By decomposing $\delta \phi^i$ into the scalar (longitudinal) type and the vector (transverse) type
\begin{align}
\delta \phi_i = \partial_i \delta \phi + \delta \phi_i^V
\,, \qquad \partial_i  \delta \phi_i^V =0
\,,
\end{align}
the quadratic action of $\delta \phi_i$ in the momentum space is given by
\begin{align}
\delta S^{(2)}_{\rm m} &=\int \frac{\D t \D^3 k}{(2\pi)^3} \left[ \delta \mathcal{L}^{(2)}_S + \delta \mathcal{L}^{(2)}_V \right]
,\\
\delta \mathcal{L}^{(2)}_S &= \rho_* k^2 \left[ \left(F_h + b_S  \frac{k^2}{\Lamae^2} \right) \delta \dot{\phi}^2 - F_h c_S^2 k^2 \delta \phi^2 
+ \frac{b_{a^2}}{2\Lamae^2} \delta \ddot{\phi}^2  \right]
, \label{LdecouplingS}
\\
\delta \mathcal{L}^{(2)}_V &= \rho_* \left[ \left(F_h + b_V \frac{k^2}{\Lamae^2} \right) (\delta \dot{\phi}^V_i)^2
-F_h c_V^2 k^2  (\delta \phi^V_i)^2 + \frac{b_{a^2}}{2\Lamae^2} (\delta \ddot{\phi}^V_i)^2 \right]
, \label{LdecouplingV}
\end{align}
with
\begin{align}
b_S \equiv \frac{3b_{\expansion^2}+2 b_{\sigma^2} }{6} \,, \qquad
b_V \equiv \frac{b_{\sigma^2} + b_{\omega^2}}{4}
\,,
\end{align}
and
\begin{align}
c_S^2\equiv  -\frac{3F_h + 2\bar{h} F_{hh}}{F_h} + \frac{4F_{e_2} }{3\bar{h} F_h}=  \frac{\partial \bar{p}}{\partial \bar{\rho} } + \frac{4 }{3} c_V^2
\,, \qquad
c_V^2\equiv \frac{F_{e_2}}{\bar{h} F_h} 
\,.
\end{align}

\subsubsection{\texorpdfstring{Solid and fluid: $F_h=\mathcal{O}(1)$}{Solid and fluid}}

Let us first discuss the solid phase and the fluid phase. Since the higher-derivative operators can be ignored as long as $k^2,\omega^2 \ll \Lamae^2, \Lambda^2$ where $\omega$ is the frequency, we only focus on the lowest-order terms. The dispersion relation of the scalar and vector modes are then given by
\begin{align}
\omega_X^2= c_X^2 k^2\,,
\end{align}
where $X=S,V$ denote the scalar mode and the vector mode, respectively. Both modes are propagating in the solid while the vector mode is non-propagating in the fluid phase ($c_V^2=0$).\footnote{One may think that the vanishing sound speed would signal a strong coupling. Nevertheless, $c_V^2=0$ is implied by the symmetry and it does not necessarily suggest the strong coupling~\cite{Endlich:2010hf}. However, if one starts with the solid phase and realizes the fluid phase as a limit, as discussed in Ref.~\cite{Endlich:2012pz} to justify a reheating process for the solid inflation, then the issue of strong coupling may arise. We are not interested in such a case and, therefore, we do not further discuss the strong coupling issue in the present paper.} As we have discussed in Sec.~\ref{subsec:Noether} and as examined in \cite{Dubovsky:2005xd,Endlich:2010hf}, this is a direct consequence of the invariance under $\Vdiff$. The stability conditions of the propagating modes require $F_h>0$ and $c_X^2>0$. In particular, the ghost-free condition~$F_h>0$ leads to the null energy condition of the background stress-energy tensor~$w>-1$.

\subsubsection{\texorpdfstring{Non-ideal aether: $F_h \ll 1 $}{Non-ideal aether}}
We then consider the non-ideal aether which is a phase of continuum satisfying $F_h \ll 1$. In this case, the higher-derivative corrections~$k^2 \delta \dot{\phi}^2, k^2 (\delta \dot{\phi}^V_i)^2$ dominate over the standard kinetic terms~$\delta \dot{\phi}^2,(\delta \dot{\phi}^V_i)^2$ for the modes with
\begin{align}
\frac{F_h \Lamae^2}{b_X} \ll k^2 \ll \Lamae^2~(\ll \Lambda^2)
\,.
\end{align}
The Nambu-Goldstone bosons of the non-ideal aether must show peculiar behaviors, stemming from the new leading operators. Here, we assume $F_h>0$, $b_X>0$, and $c_X^2>0$ to make the background stable.

As the quadratic Lagrangians~\eqref{LdecouplingS} and \eqref{LdecouplingV} contain the ghostly higher-order time derivatives, we should first understand whether the non-ideal aether is a well-defined low-energy EFT. 
The equations of motion are fourth-order differential equations and the dispersion relations are obtained from
\begin{equation}
\omega^4+\left( \mugh^2 + \cgh^2 k^2 \right)\omega^2-\mugh^2c_X^2k^2=0\,,
\end{equation}
where we have defined
\begin{align}\label{mu2}
\mugh^2 \equiv \frac{2F_h }{b_{a^2}} \Lamae^2 \,,
\qquad \cgh^2 \equiv \frac{2b_X }{b_{a^2} } \,.
\end{align}
Among the four solutions for $\omega$, two correspond to the physical modes and the other two correspond to the ghost modes, according to the Ostrogradsky theorem.
The dispersion relation of the physical modes is given by
\begin{align}
\omega^2_X&= \frac{1}{2} \left( \mugh^2 + \cgh^2 k^2 \right) \left[ -1 +\sqrt{1+\frac{4\mugh^2 c_X^2 k^2}{(\mugh^2+\cgh^2 k^2)^2} } \right]
\nn
&=\frac{\mugh^2 c_X^2 k^2}{\mugh^2+\cgh^2 k^2} \left[ 1 - \frac{\mugh^2 c_X^2 k^2}{(\mugh^2+\cgh^2 k^2)^2} + \cdots \right]\,,
\label{omega_physical}
\end{align}
while the dispersion relation of the ghosts is 
\begin{align}
\omega^2_{{\rm gh},X}&=\frac{1}{2} \left( \mugh^2 + \cgh^2 k^2 \right) \left[ -1 - \sqrt{1+\frac{4\mugh^2 c_X^2 k^2}{(\mugh^2+\cgh^2 k^2)^2} } \right]
\nn
&=-\mugh^2 - \cgh^2 k^2 + \cdots
\,.
\label{omega_ghost}
\end{align}
The scale~$\mugh^2 $ determines the mass scale at which the ghost modes with $\omega^2 \ll |\mugh^2|$ (such that $|\mugh^2|\ll\Lamae^2$) can be safely ignored (i.e.~integrated out); that is why the ghosts can be harmless within the context of the EFT. In particular, the condition~$\omega^2_X \ll |\mugh^2|$ is guaranteed by
\begin{align}
\frac{|b_{a^2}|c_X^2}{b_X} \ll 1
\,,
\label{cond_nonidealae}
\end{align}
under which we also find
\begin{align}
\frac{|\mugh^2| c_X^2 k^2}{(\mugh^2+\cgh^2k^2)^2} \ll 1
\,.
\end{align}
Hence, we can safely ignore the ghostly operator~$a_i a^i$ under the condition~\eqref{cond_nonidealae}. It is worthwhile mentioning that the operator~$a_i a^i$ can be ignored even for $b_X,b_{a^2}=\mathcal{O}(1)$ if $c_X^2 \ll 1$. In the case of small sound speed, the scaling of the space and the time is anisotropic which ensures that the higher-order time-derivative operators are irrelevant as with the ghost condensation~\cite{Arkani-Hamed:2003pdi,Arkani-Hamed:2003juy}.\footnote{When $c_X^2 \ll 1$ such that $c_X^2\Lambda^2 \ll \Lamae^2$, higher-order terms of the solid and fluid such as $\frac{1}{\Lambda^2} \spatiald_i n \spatiald^i n$ have to be taken into account in $c_X^2 \Lambda^2\ll  k^2 \ll \Lamae^2$. These effects may be implemented by replacing $c_X^2$ according to $c_X^2 \to c_X^2 + \alpha_X k^2/\Lambda^2$ with a constant~$\alpha_X$.}

Therefore, we only take into account $\expansion^2, \sigma_{ij}\sigma^{ij}$, and $\omega_{ij}\omega^{ij}$ as the leading higher-derivative operators of the non-ideal aether. The dispersion relation of the non-ideal aether is approximated as
\begin{align}
\omega_X^2 = \frac{\mugh^2 c_X^2 k^2}{\mugh^2+\cgh^2k^2} =
\begin{cases}
c_X^2 k^2 &(k^2 \ll F_h \Lamae^2) \\
F_h \Lamae^2 c_X^2/b_X  & (F_h \Lamae^2/b_X \ll k^2 \ll \Lamae^2 )\,.
\end{cases}
\end{align}
The Nambu-Goldstone bosons are certainly massless but they mimic the massive dispersion relation $\omega^2_X = $ constant for the high-frequency modes with $F_h \Lamae^2/b_X \ll k^2 \ll \Lamae^2$.

\subsubsection{\texorpdfstring{Integrable aether: $F_h\to 0, b_{a^2}\to 0$ with $\mugh^2$ kept finite}{Integrable aether}}

As we can see from Eq.~\eqref{mu2}, the ghosts become massless when one considers the limit~$F_h \to 0$ while keeping $b_{a^2}$ finite, which would signal the breakdown of the EFT. Small but non-vanishing $F_h$ would be required to justify the EFT unless $b_{a^2}=0$ is protected against quantum corrections for some reason. Nonetheless, we may consider the limit~$F_h \to 0$ of the non-ideal aether by simultaneously taking $b_{a^2} \to 0$ so that the ghost mass, determined by $\mugh^2$, is kept finite. We shall call this continuum the integrable aether because the dynamics of the Nambu-Goldstone bosons are integrable thanks to the high symmetry.

In the limit~$F_h \to 0$, both scalar and vector modes are non-propagating ($\omega^2_X =0$) due to the absence of the gradient terms. We have already seen a similar behavior in the fluid: the vector modes of the fluid are non-propagating due to the $\Vdiff$ invariance. The aether is a more symmetric object and its property is unchanged even under compression or dilatation. Accordingly, not only the vector modes but also the scalar mode does not possess oscillating solutions and the dynamics of the integrable aether is completely determined by the conserved currents.

\subsubsection{\texorpdfstring{Ideal aether: $F_h=0,b_{a^2}\neq 0$}{Ideal aether}}

We finally study the ideal aether that enjoys the exact $\diff$ invariance. Let us revisit the quadratic Lagrangian under the exact $\diff$ invariance. By introducing an auxiliary variable~$\delta A$, we rewrite the quadratic Lagrangian of the scalar mode for the ideal aether as
\begin{align}
\delta \mathcal{L}_S^{(2)} = \frac{\rho_*}{\Lamae^2} k^2 \left[ \frac{1}{2}b_{a^2}\delta \dot{A}^2 + b_S k^2 \delta A^2 +\lambda ( \delta A - \delta \dot{\phi}) \right]
\,,
\end{align}
with $\lambda$ being a Lagrange multiplier.
Since exactly the same argument holds for $\delta \phi^V_i$, we shall only consider the scalar mode~$\delta \phi$ here. The corresponding Hamiltonian is
\begin{align}
\delta \mathcal{H}_S^{(2)} = \frac{1}{2} \frac{\Lamae^2}{b_{a^2} \rho_* k^2} p_A^2 - \frac{\rho_*}{\Lamae^2}k^2 \left( b_S k^2 \delta A^2 + \lambda \delta A \right) +\lambda_1(\lambda - p_{\phi}) + \lambda_2 p_{\lambda}
\,,
\end{align}
where $p_{\phi},p_A, p_{\lambda}$ are momenta conjugate to $\delta \phi, \delta A, \lambda$ and $\lambda_1, \lambda_2$ are Lagrange multipliers to impose the primary constraints, respectively. The last two terms vanish on-shell. The Hamiltonian is unbounded from below due to the linear dependency on $\lambda$ (the Ostrogradsky theorem). However, the equation of motion concludes that $\lambda$ is a constant of motion and then the Hamiltonian on the trajectory of a solution can be bounded from below. We stress that the existence of the constant of motion is not accidental; this is a consequence of the $\diff$ symmetry as we discussed in Sec.~\ref{subsec:Noether}. As long as the $\diff$ symmetry is exact, we may get rid of the unboundedness of the Hamiltonian by restricting our consideration to solutions satisfying a given initial condition even if $b_{a^2}\neq 0$.

We emphasize the difference between the non-ideal aether with $b_{a^2}\neq 0$ and the ideal aether. In the non-ideal aether, we have the gapless physical modes~$\omega^2_X$ and the gapped ghostly modes~$\omega^2_{{\rm gh},X}$, the latter of which do not appear in the regime of validity of the EFT under the condition \eqref{cond_nonidealae}. The modes~$\omega^2_X$ become non-propagating under the ideal limit. In the ideal aether, we instead regard~$\omega_X^2$ as the ghostly global modes while the originally ghostly modes~$\omega^2_{{\rm gh},X}$ are interpreted as the gapless propagating physical degrees of freedom. The modes~$\omega_X^2$ are ghost but they are never created during the time evolution thanks to the $\diff$ symmetry. Any small violation of the $\diff$ invariance renders the theory inconsistent. The ideal aether is an object disconnected from other phases of the continuum.

The Einstein-aether theory arises as a special case of the ideal aether. As shown in Appendix~\ref{app:Einsteinaether}, there is a field transformation from the triplet of scalars~$\phi^i$ in our setup to a vector field~$A_{\mu}$ with the constraint~$A_{\mu}A^{\mu}+1=0$ by assuming a global property for the spacetime. The equation of motion for $A_{\mu}$ is a second-order differential equation even if $b_{a^2}\neq 0$ and thus free from the Ostrogradsky ghost.

\section{Gravitating continuum}
\label{sec:gravitating_continuum}

In this section, we discuss the dynamics of the continuum with self-gravity and perform the full analysis of the perturbations in the unitary gauge~$\delta \phi^i = 0 $ (the Lagrangian frame). All information of the continuum is embedded in the spacetime metric~$U,U_i,h_{ij}$; the kinematical quantities of the continuum are computed by the metric as summarized in Table~\ref{table:notation}, and the energy density, the energy flux, and the stress tensor are read by the variations of the matter action with respect to $U$, $U_i$ and $h_{ij}$, respectively.

\subsection{Expansion around gravitating background}
\label{sec:gravitatingBG}
The background should be time-dependent due to the gravitational attraction. The homogeneous and isotropic background is given by the flat FLRW metric with
\begin{align}
\langle U \rangle= \bar{U}(t)\,, \qquad \langle U_i \rangle =0\,, \qquad \langle h_{ij} \rangle = a^2(t) \delta_{ij}\,,
\end{align}
which imply
\begin{align}
\bar{h}\equiv \langle h \rangle = a^6(t)\,, \qquad \langle \expansion \rangle = 3 H(t)
\,,
\end{align}
with the Hubble expansion rate~$H\equiv \dot{a}/a = \partial_0 a/(\bar{U} a)$. We shall use a dot to denote the derivative with respect to the proper time of the background, $\dot{a}=\partial a/\partial \bar{\tau} = \partial_0 a/\bar{U}$.

We use $\alpha,\beta_i, \psi$ and $h^T_{ij}$ to parametrize the metric perturbations:
\begin{align}
U=\bar{U}(1+\alpha)\,, \qquad U_i=\beta_i/\bar{U} \,, \qquad h_{ij}=a^2 e^{2\psi} e^{h^T_{ij}}\,,
\end{align}
where $h^T_{ij}$ is traceless, $\tr h^T_{ij} = h^T_{ii}=0$.
We adopt the same notation as the decoupling-limit analysis; any index variable appearing twice in subscripts implies the summation.
The variables~$\psi$ and $h^T_{ij}$ respectively correspond to the perturbations of the determinant and the unimodular matrix
\begin{align}
h=a^6e^{6\psi}\,, \qquad \hat{h}_{ij}=e^{h^T_{ij} } 
\,.
\end{align}
As usual, the perturbations can be decomposed into the scalar-type perturbations~$\{\alpha,\beta,\psi, h^T \}$, the vector-type perturbations~$\{ \beta^V_i, h^V_i \}$, and the tensor-type perturbations~$\{ h^{TT}_{ij} \}$ via
\begin{align}
\beta_i=\partial_i \beta + \beta^V_i\,, \qquad
h^T_{ij} &=\left( \partial_i \partial_j - \frac{1}{3} \delta_{ij} \partial^2 \right) h^T + 2 \partial_{(i} h^V_{j)} + h^{TT}_{ij}\,,
\end{align}
where the vector-type is transverse and the tensor-type is transverse-traceless, respectively:
\begin{align}
\partial_i \beta^V_i = \partial_i h^V_i =0
\,, \qquad
\partial_i h^{TT}_{ij}=0\,, \qquad h^{TT}_{ii}=0
\,.
\end{align}
Although the vector and tensor perturbations can be further decomposed into polarization eigenstates, we do not explicitly decompose them because the parity symmetry is preserved in our EFT. The different types of perturbations are decoupled at the quadratic order in the action (at the linear order in the equations of motion). The dynamics of the perturbations will be discussed in the next subsection.

All types of the continua enjoy the time diff invariance
\begin{align}
t \to t'=t +  \xi_{\parallel}(t,\bm{x})/\bar{U}
\,.
\end{align}
The corresponding infinitesimal transformations of the perturbation variables are given by
\begin{align}
\alpha \to \alpha - \dot{\xi}_{\parallel}\,, \qquad
\beta_i \to \beta_i + \partial_i \xi_{\parallel} \,, \qquad
\psi \to \psi - H \xi_{\parallel}
\,, \qquad
h^T_{ij} \to h^T_{ij}
\,.
\end{align}
This shows that one of the variables is a gauge mode and we can obtain gauge-invariant variables~$\{\Phi, \mathcal{R}, \beta^V_i, h^T, h^V_i, h^{TT}_{ij}\}$ where 
\begin{align}
\Phi\equiv \alpha+\dot{\beta}\,, \qquad \mathcal{R} \equiv \psi - \frac{1}{6}\partial^2 h^T + H \beta
\,.
\end{align}
The physical meaning of the gauge-invariant variables are clarified by computing the curvature and the kinematical quantities:
\begin{align}
    \spatialR &= -4\frac{\partial^2}{a^2}\mathcal{R} +\cdots
    \,, \\
    \sigma_{ij} &=  \frac{1}{2} a^2 \dot{h}^T_{ij} +\cdots
    \,, \\
    \omega_{ij} &= \partial_{[i} \beta^V_{j]} + \cdots
    \,, \\
    a_i & = \partial_i\Phi + \dot{\beta}_i^V + \cdots\,,
\end{align}
where the ellipses are terms higher-order in perturbations. Therefore, $\mathcal{R}$ is the curvature perturbation, $h^T_{ij}=\{h^T,h^V_i,h^{TT}_{ij}\}$ are the shear perturbations, $\beta^V_i$ are the vorticity perturbations, and $\Phi$ is the perturbation of the gravitational potential in the Lagrangian frame, respectively.

Let us expand the Lagrangian in terms of the perturbation variables. Recall that we have two expansion parameters, i.e., the size of derivatives (energy and momenta) in the unit of the cutoff scale and the amplitude of perturbations around the background. Although the expansion scalar~$\expansion$ (and its time derivatives) has the background value, we can use $\expansion$ as an expansion parameter of the derivative expansion.\footnote{\label{footnote:expansion} One can split it into $\expansion=3H(t)+\delta \expansion$ and expand the action in terms of $\delta \expansion$ as performed in the EFT of inflation/dark energy~\cite{Creminelli:2006xe,Cheung:2007st,Creminelli:2008wc,Gubitosi:2012hu,Bloomfield:2012ff,Gleyzes:2013ooa,Gleyzes:2014rba} and the EFT of vector-tensor theories~\cite{Aoki:2021wew}. In this case, one should impose the consistency relations between the expansion coefficients so that the action respects the time diffs as a whole. See~\cite{Aoki:2021wew} for a detailed discussion.} In contrast, the determinant~$h={\rm det}h_{ij}$ is of order unity at the background. The perturbation~$\delta h=h-\bar{h}(t)$ is not invariant under the time diffs, differently from the unimodular part~$\delta \hat{h}_{ij}$. The easiest way to deal with the determinant~$h$ is to adopt the uniform number density slice
\begin{align}
\psi = 0 \quad \Leftrightarrow \quad n=h^{-1/2}=a^{-3}(t)\,,
\end{align}
by use of the freedom of the time diffs, provided $H\neq 0$. The solid/fluid part of the Lagrangian is then given by
\begin{align}
F(h,e_2,e_3)=F_0(h) + F_{e_2}(h) e_2 + \cdots\,,
\end{align}
up to the quadratic order in perturbations where $F_0$ and $F_{e_2}$ are independent functions of $h=a^6(t)$ and then implicit functions of time.

In summary, our gauge conditions are the uniform number density slice~$n=n(t)=a^{-3}(t)$ combined with the unitary/comoving condition~$\delta \phi^i=0=u^i$. These gauge conditions completely fix the spacetime diffs except the freedom of the time reparametrization, $t\to t'(t)$. The leading-order action perturbed around the background is then given by
\begin{align}
S[\alpha,\beta_i,h^T_{ij}]&= \int \D t \D^3 {\bm x} U(t,\bm{x}) a^3(t) \mathcal{L}
\nn
\mathcal{L}&= \frac{\Mpl^2}{2}R[g]+ \rho_* \left[ F_0 + F_{e_2} e_2+ \frac{1}{2\Lamae^2}\left( b_{\expansion^2} \expansion^2 + b_{\sigma^2}\sigma_{ij}\sigma^{ij} + b_{\omega^2} \omega_{ij}\omega^{ij} \right) \right]
.
\label{perturbed_action}
\end{align}
The functions~$F_0(t)$ and $F_{e_2}(t)$ do not involve the perturbations. We ignore the operator~$a^i a_i$ in the following analysis since this operator should be irrelevant except for the ideal aether. We will separately study the effect of $a^i a_i$ in Appendix~\ref{app:ideal_ae}.

Although we only focus on the quadratic action in the present paper, one can easily extend the action to higher orders in both perturbations and derivatives. In particular, the action of the perfect fluid coupled to GR is given by
\begin{align}
S_{\text{GR+perfect fluid}}= \int \D t \D^3 {\bm x} U(t,\bm{x}) a^3(t)  \left[ \frac{\Mpl^2}{2}R[g] - \rho(t) \right]
,
\label{GR+perfectfluid}
\end{align}
in our gauge choice where $\rho=-\rho_* F_0$ is the energy density of the perfect fluid. Note that the uniform number density slice~$n=n(t)$ agrees with the uniform density slice~$\rho=\rho(t)$ in the case of the perfect fluid. At the lowest order in derivative, the operator of the fluid is unique whereas the solid has one or more than one independent operators at each order in perturbations. The cubic, quartic, and quintic interactions of the solid are respectively given by $e_3, e_2^2$, and $ e_2 e_3$ with time-dependent coefficients and there are two operators~$e_2^3$ and $e_3^2$ at the sextic order.

We briefly mention that our EFT may be interpreted as a Lorentz-violating theory of modified gravity with the unimodular spatial metric. The modification of the Einstein-Hilbert action is transparent by writing the Lagrangian in the form
\begin{align}
\mathcal{L}=\frac{\Mpl^2}{2}\Biggl[ & \spatialR + (1+\alpha_{\sigma^2})\sigma_{ij} \sigma^{ij}-\frac{2}{3}(1+\alpha_{\expansion^2})\expansion^2  +(1+\alpha_{\omega^2}) \omega_{ij}\omega^{ij} 
\nn
& - 2\Lambda_g(t) - \frac{m_g^2(t)}{4}\left( [\delta \hat{h}^2]  - [\delta \hat{h}]^2 \right) \Biggl]\,,
\label{MGLagrangian}
\end{align}
where
\begin{align}
\alpha_{\sigma^2}\equiv \frac{\rho_* }{\Mpl^2 \Lamae^2} b_{\sigma^2}  \,, \qquad
 \alpha_{\expansion^2} \equiv  -\frac{3\rho_* }{2\Mpl^2 \Lamae^2} b_{\expansion^2} \,, \qquad 
\alpha_{\omega^2} \equiv \frac{\rho_*  }{\Mpl^2 \Lamae^2} b_{\omega^2} 
\,,
\end{align}
and
\begin{align}
\Lambda_g(t) \equiv  - \frac{\rho_* }{\Mpl^2} F_0 \,, \qquad m_g^2(t) \equiv \frac{4\rho_*}{\Mpl^2}  F_{e_2} 
\,.
\end{align}
The operators of the aether modify the kinetic terms of the graviton while the operators of the fluid and the solid yield a time-dependent ``cosmological constant'' and a time-dependent ``mass'' of the graviton, respectively.\footnote{Although the notion of mass is ambiguous in the time-dependent background, we shall simply refer to the coefficient of the quadratic term as the mass. }
For convenience, we will use the combinations~$\{ \alpha_{\sigma^2},\alpha_{\expansion^2},\alpha_{\omega^2} \}$ instead of $\{ b_{\sigma^2}, b_{\expansion^2}, b_{\omega^2} \}$ in the following discussions. Note that $\{ \alpha_{\sigma^2}, \alpha_{\expansion^2}, \alpha_{\omega^2} \}$ would be small quantities; for instance, assuming $F_0 = \mathcal{O}(1)$ and using the Friedmann equation~\eqref{Fri_eq} derived below, we obtain the relations
\begin{align}
b_{\sigma^2},b_{\expansion^2},b_{\omega^2} = \mathcal{O}(1) 
\quad \implies \quad
\alpha_{\sigma^2},\alpha_{\expansion^2},\alpha_{\omega^2} = \mathcal{O}(H^2/\Lamae^2)
\,,
\label{naive_scaling}
\end{align}
implying that the modification of the Einstein-Hilbert action is suppressed by the cutoff~$\Lamae$.\footnote{Here, a key assumption is $F_0 = \mathcal{O}(1)$. If we accept a fine-tuning~$F_0 \ll 1$ similarly to the cosmological constant problem, we can have an order-unity modification of GR for $\rho_*={\cal O}(\Mpl^4)$ and $\Lamae={\cal O}(\Mpl)$. }

The background equations of motion are obtained by imposing the absence of tadpole terms. The action is
\begin{align}
S=\int &\D \bar{\tau} \D^3 {\bm x} a^3 \Biggl\{  \bar{\mathcal{L}}(t)  + \alpha \left[ \rho_* F_0 + 3\Mpl^2\left(1+ \alpha_{\expansion^2} \right)H^2 \right] 
+ \delta \mathcal{L}^{(2)}[\alpha,\beta_i,h^T_{ij}] \Biggl\}\,,
\end{align}
where $\bar{\mathcal{L}}$ is the background part and $\delta \mathcal{L}^{(2)}$ starts at the quadratic order in perturbations. Hence, the absence of the tadpole term yields the Friedmann equation
\begin{align}
3\Mpl^2 H^2 =  \bar{\rho}\,, 
\label{Fri_eq}
\end{align}
with the energy density of the background\footnote{Alternatively, one may introduce a cosmological gravitational constant~$M_{\rm cosm}^2\equiv (1+\alpha_{\expansion})\Mpl^2$ by which the Friedmann equation is given by $3M_{\rm cosm}^2H^2=-\rho_* F_0$. }
\begin{align}
\bar{\rho}=-\rho_* F_0 -3\alpha_{\expansion^2}  \Mpl^2 H^2
= -\rho_* F_0 + \mathcal{O}(\Lamae^{-2})
\,.
\end{align}
As long as the EFT cutoff scale~$\Lamae$ is much larger than the inverse of the background time scale~$H$, which should hold to justify the use of the EFT, we can safely ignore the higher-derivative correction to the background. The spatial components of the Einstein equation are obtained by taking the derivative of \eqref{Fri_eq} as a consequence of the time diff invariance. Using the chain rule
\begin{align}
\frac{\partial F_0}{\partial \bar{\tau} } = \frac{\partial h}{\partial \bar{\tau} } \frac{\partial F_0}{\partial h }
=6 H h F_{h}\,,
\end{align}
one can easily obtain the equation
\begin{align}
 \Mpl^2 \left(2\dot{H}+3H^2\right) = -\bar{p}\,,
\end{align}
where the pressure is given by
\begin{align}
\bar{p}= \rho_* (F_0 + 2hF_h) + \alpha_{\expansion^2}  \Mpl^2\left( 2\dot{H}+3H^2 \right)
\,.
\end{align}

\subsection{Perturbations}
\label{sec:perturbations}

We study perturbations around the FLRW metric. Since the sound speeds of the phonons play important roles, we again provide their expressions here:
\begin{align}
c_S^2(t) \equiv  -\frac{3F_h + 2h F_{hh}}{F_h} + \frac{4F_{e_2} }{3 h F_h}
\,, \qquad
c_V^2(t) \equiv \frac{F_{e_2}}{h F_h} 
\,,
\end{align}
where $S$ and $V$ stand for scalar and vector, respectively. We note that $hF_h$ and $h^2F_{hh}$ are represented by $H$ and its time derivatives through the background equations. In particular, the aether limit~$F_h \to 0$ leads to the de Sitter limit~$\dot{H} \to 0$ where the limit is taken to keep $c_S^2$ and $c_V^2$ finite.

\subsubsection{Tensor perturbations}
The quadratic action of the tensor perturbations is given by
\begin{align}
\delta S^{(2)}_T =  \int \frac{\D \bar{\tau} \D^3 k}{(2\pi)^3} a^3 \frac{\Mpl^2}{8}
\left[ (1+\alpha_{\sigma^2} ) (\dot{h}^{TT}_{ij})^2 - \left(\frac{k^2}{a^2} + m_g^2 \right) (h^{TT}_{ij})^2 \right]
.
\end{align}
The propagation of gravitational waves is modified by the operators of the solid and the aether. As already mentioned, the graviton has a mass stemming from the spontaneous symmetry breaking of the spatial diffs invariance. Note that the operators of the fluid and the aether do not generate mass terms because the $\Vdiff$ invariance prohibits mass terms of the unimodular part of the ortho-spatial metric~$\hat{h}_{ij}$. We emphasize that the graviton mass is proportional to the sound speed of the transverse phonon
\begin{align}
m_g^2 = 4 (1+\alpha_{\expansion^2}) \epsilon c_V^2 H^2\,,
\end{align}
where $\epsilon = -\dot{H}/H^2$ is the slow-roll parameter. We will see similar properties in the vector and scalar perturbations below.

The operator~$\sigma_{ij}\sigma^{ij}$ changes the speed of gravitational waves. If the current universe is filled with our continuum, say dark energy is described by the aether, the observations of gravitational waves may put a constraint on the propagation speed, namely $|\alpha_{\sigma^2}| \lesssim 10^{-15}$~\cite{LIGOScientific:2017zic}. Assuming the power counting~\eqref{naive_scaling}, this constraint reads
\begin{align}
\Lamae \gtrsim 10^{-26}~{\rm eV} \sim 1~{\rm kpc}^{-1}
\,.
\end{align}
However, the observed frequency of the gravitational wave is about
\begin{align}
100~{\rm Hz} \sim 10^{-13}~{\rm eV}
\,,
\end{align}
implying that the EFT with $|\alpha_{\sigma^2}| \sim 10^{-15}$ is beyond the regime of validity in the observed frequency. Hence, the constraint~$|\alpha_{\sigma^2}| \lesssim 10^{-15}$ may not be directly applied to our EFT. See~\cite{deRham:2018red} for a related discussion.

\subsubsection{Vector perturbations}
The quadratic action of the vector perturbations takes the form
\begin{align}
\delta S^{(2)}_V[h^V_i,\beta^V_i] = \int \frac{\D \bar{\tau} \D^3 k}{(2\pi)^3} a \frac{\Mpl^2}{2}
\Biggl[ & \frac{k^2}{2}(1+\alpha_{\sigma^2}) (a \dot{h}^V_i)^2 - k^2 \beta^V_i \dot{h}^V_i 
+\left( \frac{1+\alpha_{\omega^2} }{2} \frac{k^2}{a^2} -2\dot{H} \right) (\beta^V_i)^2 
\nn
& 
+2a^2 k^2 (1+\alpha_{\expansion^2}) c_V^2 \dot{H} (h^V_i)^2 \Biggl]\,.
\label{qL_vector}
\end{align}
The variation with respect to $\beta^V_i$ yields
\begin{align}
\beta^V_i =  \frac{k^2 \dot{h}^V_i}{(1+\alpha_{\omega^2})k^2/a^2 - 4 \dot{H} } 
\,.
\label{beta_solution}
\end{align}
Substituting this solution back into the quadratic action~\eqref{qL_vector}, we obtain the action in terms of the shear perturbation~$h^V_i$. We first focus on the leading-order part of the derivative expansion
\begin{align}
\delta S^{(2)}_V[h^V_i]=\int \frac{\D \bar{\tau} \D^3 k}{(2\pi)^3} a^3 \Mpl^2 \left[ \frac{k^2}{4-k^2/(a^2\dot{H})} (\dot{h}^V_i)^2 + c_V^2 k^2 \dot{H} (h^V_i)^2 +\mathcal{O}(\Lamae^{-2})  \right]
\,,
\label{qL_vector_h}
\end{align}
which agrees with the quadratic action shown in \cite{Endlich:2012pz}. The action has a non-local kinetic term. This non-locality can be removed by performing a canonical transformation. Following Appendix~B of \cite{DeFelice:2015moy}, we reintroduce (integrate in) the vorticity perturbation~$\beta^V_i$ and consider a deformation of the action~\eqref{qL_vector_h}:
\begin{align}
\delta \tilde{S}^{(2)}_V[\beta^V_i,h^V_i]=\delta S^{(2)}_V[h^V_i] + \int \frac{\D \bar{\tau} \D^3 k}{(2\pi)^3} a^3 A(t) \left(\beta^V_i - \frac{k^2 \dot{h}^V_i}{(1+\alpha_{\omega}^2)k^2/a^2 - 4 \dot{H} }\right)^2 
\,,
\label{qL_vector2}
\end{align}
with a function~$A(t)$. The two actions~$\delta S^{(2)}_V[h^V_i]$ and $\delta \tilde{S}^{(2)}_V[\beta^V_i,h^V_i]$ are on-shell equivalent; the variation of \eqref{qL_vector2} with respect to $\beta^V_i$ yields the same solution~\eqref{beta_solution} and the same action~\eqref{qL_vector_h} is obtained by integrating out $\beta^V_i$. Let us choose $A$ so that \eqref{qL_vector2} does not have $(\dot{h}^V_i)^2$. Then, the variation of \eqref{qL_vector2} with respect to $h^V_i$ determines $h^V_i$ in terms of $\beta^V_i$ and its derivative as long as $\dot{H}\neq 0$. Thus, we obtain the quadratic action in terms of the vorticity perturbation:
\begin{align}
\delta \tilde{S}^{(2)}_V[\beta^V_i]=\int \frac{\D \bar{\tau} \D^3 k}{(2\pi)^3} a^3 \Mpl^2 \frac{ \epsilon H^2}{c_V^2 k^2} \left[ (\dot{\beta}^V_i - 3 c_S^2 H \beta^V_i)^2 - c_V^2\left( \frac{k^2}{a^2}+12 c_S^2 H^2 \right) (\beta^V_i)^2 +\mathcal{O}(\Lamae^{-2})  \right]
\,.
\label{qL_vector3}
\end{align}
The action~\eqref{qL_vector3} now has a local kinetic term. We also stress that the vorticity perturbation is massive in the presence of gravity. The mass is proportional to the sound speed of the scalar sector~$c_S^2$.

The higher-derivative corrections cannot be neglected in the aether limit which corresponds to the de Sitter limit of the background spacetime~$\dot{H} \to 0$. The quadratic action is given by
\begin{align}
\delta S^{(2)}_V[h^V_i] \to \int \frac{\D \bar{\tau} \D^3 k}{(2\pi)^3} a^3 \frac{\Mpl^2 k^2}{4}
\left( \alpha_{\sigma^2} + \frac{\alpha_{\omega^2} }{1+\alpha_{\omega^2} } \right) (\dot{h}^V_i)^2 
\,,
\end{align}
in the de Sitter limit~$\dot{H} \to 0$. We should impose $\alpha_{\sigma^2} + \alpha_{\omega^2}/(1+\alpha_{\omega^2}) = \alpha_{\sigma^2}+\alpha_{\omega^2} + \mathcal{O}(\Lamae^{-2}) \propto b_V + \mathcal{O}(\Lamae^{-2})  > 0$ to have a continuous aether limit from the solid/fluid phase (the kinetic term has to have a positive coefficient). The equation of motion of $h^V_i$ leads to $\dot{h}^V_i \propto a^{-3}$, meaning that the perturbations decay as the volume expands.

\subsubsection{Scalar perturbations}
Qualitative behaviors of the scalar perturbations are similar to those of the vector perturbations. The variables~$\alpha$ and $\beta$ are non-dynamical and can be integrated out in the absence of the operator~$a_i a^i$, yielding
\begin{align}
\delta S^{(2)}_S[h^T] = \int \frac{\D \bar{\tau} \D^3 k}{(2\pi)^3} a^3 \frac{\Mpl^2 k^4}{4}\left[ \frac{1}{3-k^2/(a^2 \dot{H})} \left( \dot{h}^T -\frac{\dot{H}}{H} h^T \right)^2 + c_S^2 \dot{H} (h^T)^2 +\mathcal{O}(\Lamae^{-2}) \right] \,. 
\label{qL_scalar_h}
\end{align}
This action coincides with that in \cite{Endlich:2012pz}.\footnote{Our gauge choice is different from the one in \cite{Endlich:2012pz}. Our variables and their variables are related by $h^T_{\rm ours} =-2 \pi_{L,{\rm theirs}}/k=6\zeta_{\rm theirs}/k^2$ and $\mathcal{R}_{\rm ours}=\mathcal{R}_{\rm theirs}$.} The relation between the curvature perturbation~$\mathcal{R}$ and the shear perturbation~$h^T$ is given by
\begin{align}
\mathcal{R} = a^2 \frac{H \dot{h}^T - \dot{H} h^T}{2(1-3a^2\dot{H}/k^2 )} +\mathcal{O}(\Lamae^{-2})  
\,.
\end{align}
Performing a canonical transformation (see e.g.~Appendix~B of \cite{DeFelice:2015moy}), one can find the quadratic action in terms of the curvature perturbation:
\begin{align}
\delta \tilde{S}^{(2)}_S[\mathcal{R} ] = \int \frac{\D \bar{\tau} \D^3 k}{(2\pi)^3} a^3 \Mpl^2 \frac{\epsilon}{c_S^2} \left[ \left( \dot{\mathcal{R}} + 4c_V^2 H \mathcal{R} \right)^2 - c_S^2 \left( \frac{k^2}{a^2} + 12c_V^2 H^2 \right) \mathcal{R}^2  +\mathcal{O}(\Lamae^{-2})  \right]\,.
\end{align}
The curvature perturbation is massive when the sound speed of the transverse phonons is non-vanishing.
Therefore, the scalar, vector, and tensor modes all acquire masses in the gravitating solid. Their masses are related to the sound speed of the different propagation states. The appearance of the mass of the curvature perturbation gives a simple explanation as to why the super-horizon curvature perturbation is not conserved in the solid inflation.

The dynamics is integrable in the aether limit. The action~\eqref{qL_scalar_h} becomes
\begin{align}
\delta S^{(2)}_V[h^T] \to  \int \frac{\D \bar{\tau} \D^3 k}{(2\pi)^3} a^3 \frac{\Mpl^2 k^4}{12}
\left( \alpha_{\sigma^2} - \alpha_{\expansion^2} \right) (\dot{h}^T)^2 
,
\end{align}
which leads to $\dot{h}^T \propto a^{-3}$. Note that $\alpha_{\sigma^2}-\alpha_{\expansion^2} \propto b_S>0$ is required to take a smooth limit. Hence, both scalar and vector modes of the shear perturbations decay with the same law in the aether phase.

\section{Discussions}
\label{sec:discussion}

Let us conclude this paper by discussing several aspects and mentioning possible future directions.

\begin{itemize}

\item {\it Higher-derivative corrections and dissipations.} Higher-derivative corrections discussed in the present paper should be distinguished from dissipative effects and a possible extension of the EFT is the inclusion of dissipations as studied in~\cite{Kovtun:2014hpa,Haehl:2015pja,Harder:2015nxa,Crossley:2015evo,Haehl:2015uoc,Jensen:2017kzi,Glorioso:2017fpd,Haehl:2018lcu}. Yet, it would be intriguing to discuss the non-dissipative higher-derivative corrections because there may be a situation where dissipations can be negligible such as cosmology mentioned below. Since we have clarified the essential building blocks as summarized in Table~\ref{table:blocks} and the variables~$\{U,U_i,h_{ij} \}$ are conjugate to the energy~$\rho$, the flux~$q^i$ and the stress tensor~$\tau^{ij}$, one may systematically find higher-derivative corrections and discuss how they affect $\{\rho,q^i,\tau^{ij} \}$.

\item {\it Solid inflation.} Solid inflation~\cite{Endlich:2012pz} is a model of the cosmic inflation with peculiar features stemming from the symmetry breaking pattern for the solid. As it is well known, the consistency conditions among correlation functions for single-field inflation are violated in solid inflation~\cite{Endlich:2013jia,Dimastrogiovanni:2014ina,Akhshik:2014bla,Bordin:2016ruc}, and our EFT formulation may be used to extract universal features of non-Gaussianities in solid inflation. For instance, as mentioned in Sec.~\ref{sec:gravitatingBG}, interactions of the solid at the zeroth order in derivative are given by polynomials of $e_2$ and $e_3$ which involve both scalar and tensor perturbations. A new interaction of the solid should yield new interactions of both scalar and tensor perturbations, suggesting the existence of consistency relations between amplitudes of scalar and tensor non-Gaussianities. On the other hand, in the EFT of single-field inflation~\cite{Cheung:2007st}, the operators analogous to $e_2^p e_3^q$ are $(g^{00}+1)^n$ that only involve scalar perturbations. Thus, the EFTs can reveal the different structures of nonlinear interactions stemming from the underlying symmetry breaking patterns. Moreover, our EFT formulation in the unitary gauge will be more useful to investigate a reheating scenario for solid inflation.

\item {\it Aether as dark energy.} The late-time cosmic acceleration (dark energy) is one of the major mysteries of modern cosmology and our aether may be a natural candidate for dark energy. Indeed, when a continuum approaches the equation of state with $w= -1$, the symmetry is enhanced and then the operators of aether have to be taken into account. The aether is almost a cosmological constant, but it has perturbations which may yield observational features different from the standard $\Lambda$CDM cosmology. In the light of observational tensions of the $\Lambda$CDM cosmology~\cite{Bernal:2016gxb,Planck:2018vyg}, it would be interesting to address the dark energy problem from this new direction.

\item {\it Possible extensions.} In the present paper, we assumed the global symmetry~\eqref{ISOdiag}, which is respected by the spatially flat FLRW background in the unitary gauge. It should be straightforward to extend the EFT of gravitating continuum so that the corresponding global symmetry is respected by spatially curved FLRW backgrounds as well. It is also interesting to extend the EFT to a multi-component system including interactions between different components as well as other types of materials such as framids~\cite{Nicolis:2015sra}.

\item {\it Thread-based decomposition vs.~foliation-based (ADM) decomposition.} Apart from the EFT, developing the thread-based decomposition of the spacetime is an interesting direction. The geometrical difference between the thread-based decomposition and the standard ADM decomposition is whether the spacetime is torn into timelike curves or foliated by spacelike hypersurfaces (Figure~\ref{fig:decomposition}). Since the timelike flow is the essential ingredient in relativistic hydrodynamics, the former one may provide a more appropriate framework for describing relativistic hydrodynamics and this is indeed the case for formulating the EFT. For instance, although the ADM decomposition has been used in numerical relativity, it could be more useful to adopt the thread-based decomposition when one is interested in quantities defined in the Lagrangian coordinates. Moreover, the ADM decomposition has been used to develop the Hamiltonian formulation and quantum gravity such as loop quantum gravity~\cite{Thiemann:2001gmi,Rovelli:2004tv} and Ho\v{r}ava-Lifshitz gravity~\cite{Horava:2009uw}. The thread-based and foliation-based decompositions preserve different parts of the spacetime symmetries and it might be interesting to revisit the formal aspects of gravity by means of the thread-based decomposition.

\end{itemize}

\vspace{0.7cm}

{\bf Acknowledgments:} K.A.~would like to thank Rikkyo University for their hospitality during his visit. M.A.G.~thanks Masahide Yamaguchi for discussion and the Tokyo Institute of Technology for hospitality when this work was in its final stage. The work of K.A.~was supported by JSPS KAKENHI Grant No.~19J00895, No.~20K14468 and No.~17H06359. The work of M.A.G.~was supported by MEXT KAKENHI Grant No.~17H02890. The work of S.M.~was supported in part by JSPS Grants-in-Aid for Scientific Research No.~17H02890, No.~17H06359, and by World Premier International Research Center Initiative, MEXT, Japan. The work of K.T.~was supported by JSPS KAKENHI Grant No.~JP21J00695.
\vspace{0.7cm}

\appendix

\section{Relation to the Einstein-aether theory}
\label{app:Einsteinaether}
We clarify the relation between the EFT having the generic internal diffs which we have called the aether and the Einstein-aether theory that has a fixed norm vector field rather than the triplet of scalar fields. The action of the aether generically takes the form
\begin{align}
S[g_{\mu\nu},u_\mu]&=\int \D^4x \sqrt{-g} \left[ \frac{\Mpl^2}{2}R[g_{\mu\nu}]+\Lae (g_{\mu\nu},u_{\mu},\nabla_{\mu})\right]\,, \label{S_aether} \\
u^{\mu}&\equiv -\frac{1}{6 \sqrt{{\rm det} h^{mn}} } \varepsilon_{ijk}\epsilon^{\mu\nu\rho\sigma}\partial_{\nu}\phi^i \partial_{\rho}\phi^j \partial_{\sigma} \phi^k
\,,
\end{align}
where we recall that $u^{\mu}$ is invariant under the generic internal diffs. By introducing an auxiliary field~$A_{\mu}$ and Lagrange multipliers~$\lambda$ and $\lambda_i$, we can consider an equivalent action:
\begin{align}
S[g_{\mu\nu},A_\mu,\phi^i]=\int \D^4x \sqrt{-g}\left[ \frac{\Mpl^2}{2}R[g_{\mu\nu}]+ \Lae (g_{\mu\nu},A_{\mu},\nabla_{\mu}) + \lambda (A_{\mu}A^{\mu}+1) + \lambda_i A^{\mu}\partial_{\mu} \phi^i \right] \,.
\label{Seq_aether}
\end{align}
The variations with respect to $\lambda$ and $\lambda_i$ yield
\begin{align}
A_{\mu}A^{\mu}+1=0\,, \qquad A^{\mu}\partial_{\mu} \phi^i =0\,,
\end{align}
which are solved by 
\begin{align}
A^{\mu}=u^{\mu} =  -\frac{1}{6 \sqrt{{\rm det} h^{mn}} } \varepsilon_{ijk}\epsilon^{\mu\nu\rho\sigma}\partial_{\nu}\phi^i \partial_{\rho}\phi^j \partial_{\sigma} \phi^k
\,.
\end{align}
We then recover the original aether action~\eqref{S_aether} by substituting this solution into the action~\eqref{Seq_aether}. We instead take the variation of \eqref{Seq_aether} with respect to $\phi^i$:
\begin{align}
\nabla_{\mu}(\lambda_i A^{\mu}) = 0
\,,
\label{lameom}
\end{align}
of which the solution is given by
\begin{align}
\lambda_i A^{\mu} = \epsilon^{\mu\nu\rho\sigma}\partial_{\nu}B_{i \rho\sigma}\,.
\label{lamsol}
\end{align}
Here, $B_{i\rho\sigma}$ is a triplet of 2-form fields. Note that the solution~\eqref{lamsol} is unique on a contractible manifold according to the Poincar\'{e} lemma. However, this may not be generically true. For a while, we assume a contractible spacetime manifold (or we only consider local properties of the theory) and we will revisit this point below. Since $A_{\mu}$ is restricted to be $A_{\mu}A^{\mu}=-1$ due to the constraint, the solution~\eqref{lamsol} determines $\lambda_i$ and we can substitute this into the action~\eqref{Seq_aether}. We then find the action of the Einstein-aether theory,
\begin{align}
S[g_{\mu\nu},A_\mu]=\int \D^4x \sqrt{-g}\left[ \frac{\Mpl^2}{2}R[g_{\mu\nu}]+\Lae (g_{\mu\nu},A_{\mu},\nabla_{\mu}) + \lambda (A_{\mu}A^{\mu}+1) \right] \,,
\label{S_Eae}
\end{align}
concluding that the aether~\eqref{S_aether} is locally equivalent to the Einstein-aether theory~\eqref{S_Eae} when the spacetime manifold is contractible.

However, the equivalence between \eqref{S_aether} and \eqref{S_Eae} does not hold globally. Let us consider an example:
\begin{align}
\Lae(g_{\mu\nu},u_{\mu},\nabla_{\mu}) = M^2 \expansion^2
\,, \qquad
\Lae(g_{\mu\nu},A_{\mu},\nabla_{\mu}) = M^2 (\nabla^{\mu}A_{\mu})^2
\,.
\label{example:Lae}
\end{align}
This example might be too simple but is sufficient for an illustrative purpose.
This continuum has neither the energy flux nor the anisotropic stress since the Lagrangian is independent of $U_i$ or $\hat{h}_{ij}$. The energy density and the isotropic pressure are given by
\begin{align}
\rho = M^2 \expansion^2 \,, \qquad p=-2\dertau \expansion - \expansion^2
\,,
\end{align}
and then the Euler equation~\eqref{Eulereq} reads
\begin{align}
\frac{1}{\sqrt{h}}\dertau \left( \sqrt{h} \spatiald_i \expansion \right) =0 
\,.
\label{example:eom_ae}
\end{align}
The integral of \eqref{example:eom_ae} gives
\begin{align}
\spatiald_i \expansion = \frac{\mathcal{J}_i( \bm{x} )}{\sqrt{h} }\,,
\label{example:eom_ae2}
\end{align}
where $\mathcal{J}_i( \bm{x} )$ are integration constants. On the other hand, the equations of motion derived from \eqref{S_Eae} with \eqref{example:Lae} are
\begin{align}
M^2\nabla_{\mu}\nabla^{\nu}A_{\nu}-\lambda A_{\nu}=0
\,, \qquad A_{\mu}A^{\mu}+1=0
\,,
\label{example:eom_Eae}
\end{align}
on top of the Einstein equations. Using \eqref{example:eom_Eae}, one can show 
\begin{align}
(\delta^{\mu}_{\alpha}+A^{\mu}A_{\alpha})(M^2\nabla_{\mu} \nabla^{\nu}A_{\nu}) = 0
\,,
\label{example:eom_Eae2}
\end{align}
which is equivalent to
\begin{align}
\spatiald_i \expansion =0
\,,
\label{example:eom_Eae3}
\end{align}
under the on-shell relation $A_{\mu}=u_{\mu}$. Therefore, the equations~\eqref{example:eom_ae2} and \eqref{example:eom_Eae3} are not equivalent if $\mathcal{J}_i(\bm{x})\neq 0$. The aether~\eqref{S_aether} has more freedom to choose initial conditions than \eqref{S_Eae} and the solutions of \eqref{S_aether} agree with those of \eqref{S_Eae} if and only if the initial conditions satisfy $\mathcal{J}_i(\bm{x})=0$. The same conclusion can be reached by a perturbative analysis. Let us consider the gravity decoupling limit in which the expansion scalar is given by $\expansion^2 = (\partial_i \delta \dot{\phi}^i)^2$ up to the quadratic order in perturbations. We need $3\times 2$ initial conditions to determine the dynamics of $\delta \phi^i$. Since the theory enjoys the generic internal diffs, the initial conditions on $\delta \phi^i$ have no physical consequence and only the initial conditions on the time derivatives of $\delta \phi^i$ are physical conditions. Hence, we have 3 physical initial conditions for \eqref{S_aether}. On the other hand, the gravity decoupling limit yields $(\nabla^{\mu}A_{\mu})^2 = (\partial_i \delta A^i)^2$ at the quadratic order in perturbations after solving $A_{\mu}A^{\mu}+1=0$ where $\delta A^i$ are the spatial components of the perturbation of $A^{\mu}$ around $\bar{A}^{\mu}=(1,0,0,0)$. The equations of motion of $\delta A^i$ are thus constraint equations and there is no freedom associated with the initial conditions. Hence, the number of degrees of freedom of \eqref{S_aether} is larger than that of \eqref{S_Eae}. See Sec.~\ref{subsec:decoupling} for detailed analysis on the perturbations.

The additional degrees of freedom can be thought of as global ones in the sense that their dynamics are determined by integration constants (and they are killed in a contractible manifold). As we have discussed in Sec.~\ref{subsec:Noether}, the symmetry of the aether concludes the existence of the conserved currents. The equations of motion are integrable as we have explicitly seen in \eqref{example:eom_ae2} and the integration constants~$\mathcal{J}_i(\bm{x})$ are responsible for the dynamics of the additional degrees of freedom.

\section{Curvature and torsion in the thread-based decomposition}
\label{app:1+3}

In the main text, we have used a spacetime decomposition with respect to the four-velocity of the continuum. We note that there is no unique definition of ortho-spatial curvature (and torsion) due to the absence of the spacelike hypersurface orthogonal to the four-velocity. Accordingly, several possible definitions of curvatures have been discussed. Let us briefly comment on curvature and torsion in this Appendix.

We use the holonomic basis expressions in which the ortho-spatial covariant derivative of an ortho-spatial vector~$\spatialV^{\mu}$ is given by
\begin{align}
\spatialD_{\mu} \spatialV^{\nu}
= h^{\alpha}{}_{\mu}h^{\nu}{}_{\beta} 
    \nabla_{\alpha} \spatialV^{\beta }
\,,
\end{align}
as read by \eqref{DXV}. The commutator of the covariant derivative yields
\begin{align}
2 \spatialD_{[\mu} \spatialD_{\nu]} \spatialS = 2n^{\alpha} \omega_{\mu\nu} \nabla_{\alpha} \spatialS\,,
\label{commutator_scalar}
\end{align}
for a scalar~$\spatialS$. The ortho-spatial curvature~$\spatialR^{\rho}{}_{\sigma \mu \nu}$ in the holonomic basis is computed by the commutator
\begin{align}
    2 \spatialD_{[\mu} \spatialD_{\nu]} \spatialV^{\rho}
    & = \spatialR^{\rho}{}_{\sigma \mu \nu} \spatialV^{\sigma}
    + 2\omega_{\mu\nu} h^{\rho}{}_{\beta} \pounds_u \spatialV^{\beta}
    \,, \label{commutator_vector}
\end{align}
whereas an alternative definition of the curvature is
\begin{align}
2 \spatialD_{[\mu} \spatialD_{\nu]} \spatialV^{\rho}
    & = \spatialRold^{\rho}{}_{\sigma \mu \nu} \spatialV^{\sigma}
    + 2u^{\alpha}\omega_{\mu\nu} h^{\rho}{}_{\beta} \nabla_{\alpha} \spatialV^{\beta}
    \,, \label{commutator_vector_old}
\end{align}
which has been used in the EFT of vector-tensor theories~\cite{Aoki:2021wew}. The two definitions are related by
\begin{align}
\spatialR^{\mu}{}_{\nu\rho\sigma} = \spatialRold^{\mu}{}_{\nu\rho\sigma} + 2 \spatialB_{\nu}{}^{\mu}\omega_{\rho\sigma}
\,,
\end{align}
where $\spatialB_{\mu\nu} \equiv \expansion_{\mu\nu}+\omega_{\mu\nu}$ and their symmetry properties are
\begin{align}
\spatialR_{(\mu\nu)\rho\sigma} \neq 0\,, \qquad \spatialR_{\mu\nu(\rho\sigma)}=0\,, \qquad \spatialR_{\mu[\nu\rho\sigma]} =0
\,, \\
\spatialRold_{(\mu\nu)\rho\sigma} = 0\,, \qquad \spatialRold_{\mu\nu(\rho\sigma)}=0\,, \qquad \spatialRold_{\mu[\nu\rho\sigma]} \neq 0
\,.
\end{align}
There would be no preference since both quantities are tensorial under the thread-preserving diffs and reduce to the intrinsic curvature when the vorticity vanishes. The difference between \eqref{commutator_vector} and \eqref{commutator_vector_old} is whether we use the Lie derivative~$\pounds_u$ or the parallel component of the covariant derivative~$n^{\alpha}\nabla_{\alpha}$ in the second term of the commutator. The reason to adopt \eqref{commutator_vector} as the curvature in the present paper is that not the spacetime covariant derivative~$\nabla_{\mu}$ but the Lie derivative (the material derivative) plays an important role in the EFT formulation and the expression in the unitary gauge is more analogous to the standard curvature. See \cite{Perjes:1992np,Boersma:1994pc,Roy:2014lda} for more discussions.

The commutators~\eqref{commutator_scalar} and \eqref{commutator_vector_old} are reminiscent of the commutators of a covariant derivative with a torsion. It might be reasonable to call $n^{\alpha}\omega_{\mu\nu}$ the ``torsion.'' However, as emphasized in \cite{Boersma:1994pc}, the role of $n^{\alpha}\omega_{\mu\nu}$ is different from the torsion (and thus called {\it deficiency} in \cite{Boersma:1994pc}). Indeed, the quantity~$\omega_{\mu\nu}$ represents the vorticity, characterizing ``deficiency'' of spacelike hypersurfaces orthogonal to the four-velocity. Therefore, we do not introduce the notion of the torsion in our EFT formulation although the term~``torsion'' has been used in the EFT of dissipative fluids~\cite{Crossley:2015evo}.

\section{Higher-derivative corrections to fluids and solids}\label{app:HDO}
Let us discuss higher-order derivative corrections to fluids and solids. Since the action has a freedom associated with integration by parts, we first summarize the formulae for integration by parts as follows:
\begin{align}
    \int \D^4x  \sqrt{-g}\, \dertau \spatialS &= - \int \D^4 x \sqrt{-g}\, \expansion \, \spatialS
    \,, \\
    \int \D^4x  \sqrt{-g}\, \spatialD_i \spatialV^i &=- \int \D^4 x \sqrt{-g} \, a_i \spatialV^i
    \,, \\
        \int \D^4x  \sqrt{-g}\, \spatiald_i \spatialV^i &=  \int \D^4x  \sqrt{-g}\, \hat{\spatialD}_i \spatialV^i = -\int \D^4 x \sqrt{-g} \left[ a_i + \frac{1}{2}\spatialD_i \ln h  \right] \spatialV^i
    \,, 
\end{align}
where the boundary terms are discarded. In the case of fluids/solids, the coefficients of the EFT operators are functions of $h$ (or functions of $t$ in the uniform number density slice). Then, the integration by parts yields
\begin{align}
\int \D^4x  \sqrt{-g}\, f(h) \dertau \spatialS &= - \int \D^4 x \sqrt{-g}\, \left[ f(h)  + 2h f'(h) \right] \expansion \, \spatialS
\,, \\
\int \D^4x  \sqrt{-g}\, f(h) \spatialD_i \spatialV^i &=- \int \D^4 x \sqrt{-g} \, \left[ f(h) a_i + h f'(h) \spatialD_i \ln h \right] \spatialV^i
\,.
\end{align}

The new building blocks of fluids are $h$ and its ortho-spatial covariant derivatives. Hence, the new operators at the quadratic order in derivatives are
\begin{align}
\spatialD_i h \spatialD^i h  \,, \quad a^i \spatialD_i h 
\,,
\label{fluidop_all}
\end{align}
up to the freedom of integration by parts. For instance, one can consider $\spatialD^i \spatialD_i h$ but it can be absorbed into \eqref{fluidop_all}.
Here, we recall that $n=h^{-1/2}$ represents the number density of the fluid particles. Hence, these operators contribute to the dynamics when the ortho-spatial gradient of the number density does not vanish. In particular, the gradient is given by
\begin{align}
\spatialD_i n(t) = U_i \partial_0 n(t)
\,,
\end{align}
in the uniform number density slice $n=n(t)$.

There are many new operators for solids since the new building block~$\delta \hat{h}_{ij}$ has indices. We only focus on the operators that contribute to the quadratic action. To find independent operators of solids, it is useful to use the relation
\begin{align}
\hat{h}^{ij} \spatiald_k \delta \hat{h}_{ij}=\spatiald_k \ln ({\rm det}\hat{h}_{ij}) = 0 
\,.
\end{align}
Therefore, up to the freedom of integration by parts, we find four independent second-order derivative operators of the solids which contribute in the quadratic Lagrangian of the perturbations:
\begin{align}
h^{ij} h^{kl} h^{nm} \spatiald_i \delta \hat{h}_{kn} \spatiald_j \delta \hat{h}_{lm}\,, \quad
h^{ij} \divh_i \divh_j\,, \quad
\divh_i \spatialD^i h \,, \quad
\divh_i a^i
\,,
\label{solidop_all}
\end{align}
where
\begin{align}
\divh_i \equiv h^{jk}\spatiald_j \delta \hat{h}_{ki}
\,.
\end{align}
Note that one can use $\spatialD_i \delta \hat{h}_{jk}$ alternative to $\spatiald_i \delta \hat{h}_{jk}$ to easily perform the integration by parts. Nonetheless, the difference between $\spatialD_i \delta \hat{h}_{jk}$ and $\spatiald_i \delta \hat{h}_{jk}$ appears at the quadratic order in perturbations, so it does not matter in the present discussion. The first one in \eqref{solidop_all} leads to the gradient term of the graviton which is already present in the ortho-spatial Ricci scalar~$\spatialR$. As a result, the new operators of the solid at the quadratic order in both perturbations and derivatives are
\begin{align}
h^{ij} \divh_i \divh_j\,, \quad
\divh_i \spatialD^i h \,, \quad
\divh_i a^i
\,,
\end{align}
with coefficients that are functions of $h$.

\section{\texorpdfstring{Gravitational perturbations with $a^i a_i$}{Gravitational perturbations with acceleration squared}}
\label{app:ideal_ae}
The quadratic action of the vector perturbation in the presence of the operator~$a^i a_i$, which is relevant for the ideal aether, is given by
\begin{align}
\delta S^{(2)}_V[h^V_i,\beta^V_i] = \int \frac{\D \bar{\tau} \D^3 k}{(2\pi)^3} & a \frac{\Mpl^2}{2}
\Biggl[ \alpha_{a^2} (\dot{\beta}^V_i)^2 + \frac{k^2}{2}(1+\alpha_{\sigma^2}) (a \dot{h}^V_i)^2 - k^2 \beta^V_i \dot{h}^V_i
\nn
& +\left( \frac{1+\alpha_{\omega^2} }{2} \frac{k^2}{a^2} -2\dot{H} \right) (\beta^V_i)^2 
+2a^2 k^2 (1+\alpha_{\expansion^2}) c_V^2 \dot{H} (h^V_i)^2 \Biggl]\,,
\label{qL_vector_full}
\end{align}
where $\beta^V_i$ is the vorticity perturbation and $h^V_i$ is the shear perturbation, respectively. The new term in comparison with \eqref{qL_vector} is the kinetic term of the vorticity perturbation. Thus, there are generically four propagating modes; two would be physical modes and the other two would be ghostly modes. Nonetheless, the symmetry of the aether concludes that two of the modes are integrable. Indeed, one can easily see that the equation of motion of the shear in the aether ($\dot{H}=0$) takes the form
\begin{align}
\frac{\partial}{\partial \bar{\tau} }\left[ a^3 (1+\alpha_{\sigma}) \dot{h}^V_i - a \beta^V_i \right] =0
\,.
\end{align}
After integration, the equations of motion of the vector perturbations are given by
\begin{align}
\dot{h}^V_i -\frac{\beta^V_i}{(1+\alpha_{\sigma}) a^2} &=  \frac{\mathcal{J}^V_i}{a^3}
\,, \\
2\alpha_{a^2}\left( \ddot{\beta}_i^V+H\dot{\beta}_i^V \right) - \frac{k^2}{a^2} \left( \alpha_{\omega^2} + \frac{\alpha_{\sigma^2} }{1+\alpha_{\sigma^2} } \right) \beta^V_i & = - k^2  \frac{\mathcal{J}^V_i }{a^3}
\,,
\end{align}
where $\mathcal{J}^V_i$ are integration constants. Thus, the operator~$a^i a_i$ yields two additional propagating modes sourced by the integration constants. Note that these modes should be distinguished from the transverse phonons (see Sec.~\ref{subsec:decoupling}). Moreover, the first equation roughly states that the difference between the shear and the vorticity decays as the volume expands and, therefore, the system has two propagating modes at the attractor phase in an expanding background.

We only briefly report the scalar perturbations since its qualitative behavior is the same. The quadratic action after integrating out $\alpha$ is
\begin{align}
\delta S^{(2)}_S[h^T,\mathcal{R}] = \int \frac{\D \bar{\tau} \D^3 k}{(2\pi)^3} & a^3 \Mpl^2\left[ \frac{1}{2} \dot{\chi}^t \bm{K_S} \dot{\chi} + \dot{h}^T M_S \mathcal{R} - \frac{1}{2} \chi^t \bm{V_S} \chi \right] \,, \qquad 
\chi=
\begin{pmatrix}
h^T \\ \mathcal{R}
\end{pmatrix}\,,
\label{qL_scalar_full}
\end{align}
with matrices~$\bm{K_S}, \bm{V_S} $ and a function~$M_S$. Here, we use $h^T$ and $\mathcal{R}$ as independent variables. In the presence of $a^i a_i$, the matrix~$\bm{K_S}$ is non-degenerate and then both shear and curvature perturbations are dynamical. In the aether case, the action depends only on $\dot{h}^T$, meaning that the equation of motion of $h^T$ is integrable. One can then obtain a wave equation for the curvature perturbation sourced by an integration constant.

We notice that the Hamiltonian of \eqref{qL_vector_full} can be bounded from below and above thanks to the degeneracy of gravitational Lagrangian even if $\alpha_{a^2}\neq 0$ and the $\diff$ invariance is not exact. The boundedness of the Hamiltonian of \eqref{qL_vector_full} in the high-momentum limit requires
\begin{align}
\alpha_{a^2}>0\,, \qquad 1+\alpha_{\sigma^2}>0 \,, \qquad -(1+\alpha_{\expansion^2})c_V^2\dot{H}>0
\,, \qquad 1+\alpha_{\omega^2} < 0\,.
\label{stability_vector}
\end{align}
On the other hand, the positive definiteness of $\bm{K_S}$ and $\bm{V_S}$ in high-momentum limit [the boundedness of the Hamiltonian of \eqref{qL_scalar_full}] additionally requires
\begin{align}
2>\alpha_{a^2}>0 \,, \qquad \alpha_{\sigma^2}-\alpha_{\expansion^2} > 0\,, \qquad -(1+\alpha_{\expansion^2})c_S^2\dot{H}>0 \,, \qquad 1+\alpha_{\expansion^2}<0
\,. \label{stability_scalar}
\end{align}
Note that the last inequalities of \eqref{stability_vector} and \eqref{stability_scalar} cannot hold in the gravity decoupling limit~$\Mpl^2 \to \infty$ with $\rho_*$ and $\Lamae$ kept finite, consistently with Sec.~\ref{subsec:decoupling}. Although the inequalities~\eqref{stability_vector} and \eqref{stability_scalar} may be satisfied simultaneously, it requires ``wrong'' signs of the operators~$\expansion^2$ and $\omega_{ij}\omega^{ij}$. In particular, $1+\alpha_{\expansion^2}<0$ leads to a negative cosmological gravitational constant~$M_{\rm cosm}^2=(1+\alpha_{\expansion^2})\Mpl^2<0$ which might cause some pathology, e.g.~no expanding universe for a positive energy density.

\bibliographystyle{JHEPmod}
\bibliography{refs}

\end{document}